\pdfoutput=1
\documentclass{emulateapj}
\pdfoutput=1

\usepackage{color}

\newcommand{\av}{\ensuremath{A_{V}}}
\newcommand{\kt}{\ensuremath{k_{\rm{B}}T}}
\newcommand{\lx}{\ensuremath{L_{\rm{X}}}}

\newcommand{\emm}{\ensuremath{EM}}
\newcommand{\fx}{\ensuremath{F_{\rm{X}}}}
\newcommand{\nh}{\ensuremath{N_{\rm H}}}
\newcommand{\ri}{\textit{R}}
\newcommand{\ji}{\textit{J}}
\newcommand{\hi}{\textit{H}}
\newcommand{\ki}{\textit{K}}
\newcommand{\ksi}{\textit{K\ensuremath{_{\rm{s}}}}}

\slugcomment{}
\shorttitle{High-Resolution X-ray Imaging of RCW\,49}
\shortauthors{Tsujimoto et al.}

\begin{document}
\title{An X-ray Imaging Study of the Stellar Population in RCW\,49}
\author{M.~Tsujimoto\altaffilmark{1,2,3}, E.~D.~Feigelson\altaffilmark{1},
L.~K.~Townsley\altaffilmark{1}, P.~S.~Broos\altaffilmark{1},
K.~V.~Getman\altaffilmark{1}, J.~Wang\altaffilmark{1}, G.~P.~Garmire\altaffilmark{1},
D.~Baba\altaffilmark{4}, T.~Nagayama\altaffilmark{5}, M.~Tamura\altaffilmark{6}, \&
E.~B.~Churchwell\altaffilmark{7}}

\altaffiltext{1}{Department of Astronomy \& Astrophysics, Pennsylvania State University,
525 Davey Laboratory, University Park, PA 16802.}
\altaffiltext{2}{Department of Physics, Rikkyo University, 3-34-1, Nishi-Ikebukuro,
Toshima, Tokyo 171-8501, Japan.}
\altaffiltext{3}{Chandra Fellow. E-mail: \texttt{tsujimot@astro.psu.edu}.}
\altaffiltext{4}{Department of Physics, Graduate School of Science, Nagoya University,
Furo-cho, Chikusa, Nagoya, 464-8602 Japan.}
\altaffiltext{5}{Department of Astronomy, Graduate School of Science, Kyoto University,
Kitashirakawa-Oiwake-cho, Sakyo, Kyoto, 606-8502 Japan.}
\altaffiltext{6}{National Astronomical Observatory of Japan, 2-21-1, Osawa, Mitaka,
Tokyo 181-8588, Japan.}
\altaffiltext{7}{Department of Astronomy, University of Wisconsin, 475 North Charter
Street, Madison, WI 53706.}

\begin{abstract}
 We present the results of a high-resolution X-ray imaging study of the stellar
 population in the Galactic massive star-forming region RCW\,49 and its central OB
 association Westerlund 2. We obtained a $\sim$40~ks X-ray image of a
 $\sim$17\arcmin$\times$17\arcmin\ field using the \textit{Chandra X-ray Observatory}
 and deep near-infrared (NIR) images using the Infrared Survey Facility in a concentric
 $\sim$8\farcm3$\times$8\farcm3 region. We detected 468 X-ray sources and identified
 optical, NIR, and \textit{Spitzer Space Telescope} mid-infrared (MIR) counterparts for
 379 of them. The unprecedented spatial resolution and sensitivity of the X-ray image,
 enhanced by optical and infrared imaging data, yielded the following results: (1) The
 central OB association Westerlund 2 is resolved for the first time in the X-ray
 band. X-ray emission is detected from all spectroscopically-identified early-type stars
 in this region. (2) Most ($\sim$86\%) X-ray sources with optical or infrared
 identifications are cluster members in comparison with a control field in the Galactic
 Plane. (3) A loose constraint (2--5~kpc) for the distance to RCW\,49 is derived from
 the mean X-ray luminosity of T Tauri stars. (4) The cluster X-ray population consists
 of low-mass pre--main-sequence and early-type stars as obtained from X-ray and NIR
 photometry. About 30 new OB star candidates are identified. (5) We estimate a cluster
 radius of 6\arcmin\--7\arcmin\ based on the X-ray surface number density profiles. (6)
 A large fraction ($\sim$90\%) of cluster members are identified individually using
 complimentary X-ray and MIR excess emission. (7) The brightest five X-ray sources, two
 Wolf-Rayet stars and three O stars, have hard thermal spectra.
\end{abstract}


\section{INTRODUCTION}
\ion{H}{2} regions, which are recognized by diffuse hydrogen emission ionized by UV photons
from stars earlier than B2, are the sites of massive star formation. Strong radiation
and winds from early-type stars in massive star-forming regions (SFRs) have effects that
cannot be understood by a simple extrapolation of low-mass SFRs. X-ray studies of
massive SFRs, which are potentially important to detect hot gas as well as stellar
constituents, had lagged behind those of low-mass regions in the past due to their
greater distance, source confusion, and obscuration. Both high spatial resolution and
hard X-ray sensitivity are required to overcome these difficulties.

The \textit{Chandra X-ray Observatory} is capable of conducting high-resolution and
high-sensitivity imaging observations in the 0.5--8.0~keV band. Its $\sim$1\arcsec\
spatial resolution and sensitivity are sufficient to resolve scales $\sim$10$^{17}$~cm
(a typical size of ultra-compact \ion{H}{2} regions; \citealt{churchwell02}) and to detect
X-ray emission from embedded late-type pre--main-sequence sources at a distance of
several kiloparsecs, typical of well-studied Galactic massive SFRs.

Along with the \textit{Chandra} Orion Ultradeep Project \citep{getman05}, which led to a
series of X-ray imaging studies of the nearest ($D\lesssim$~0.5~kpc) rich SFR, we have seen a
surge of results from X-ray studies of massive SFRs at larger distances. Galactic
massive SFRs with X-ray imaging surveys of their stellar contents include: NGC\,3603
\citep{moffat02}, W\,3 \citep{hofner02}, the Carina nebula
\citep{evans03,evans04,sanchawala06}, NGC\,6530 \citep{damiani04}, the Trifid Nebula
\citep{rho04}, S\,106 \citep{giardino04}, Ara OB1 \citep{skinner05}, Cepheus B
\citep{getman06}, RCW\,38 \citep{wolk06}, the Arches and Quintuplet clusters
\citep{dqwang06}, NGC\,2362 \citep{delgado06}, NGC\,2264 \citep{flaccomio06}, Westerlund
1 \citep{skinner06}, IC\,1396N \citep{getman07a}, NGC\,6357 \citep{jwang07a}, the
Rosette nebula \citep{townsley03,jwang07b}, M\,16 \citep{linsky07}, Cygnus OB2
\citep{albacete07}, and M\,17 \citep{townsley03,broos07}. Early results are reviewed in
\citet{townsley06b} and \citet{feigelson07}.

\medskip

This paper describes the results of the \textit{Chandra} imaging study of the stellar
contents in a southern massive SFR RCW\,49 and its central OB association Westerlund
2. The region was imaged in the mid-infrared (MIR) band as a part of the Galactic Legacy
Infrared Mid-Plane Survey Extraordinaire program (GLIMPSE; \citealt{benjamin03}) using
the \textit{Spitzer Space Telescope} \citep{werner04}. We also obtained deep
near-infrared (NIR) images of this region using the Infrared Survey Facility. Many
previous \textit{Chandra} studies were combined with NIR imaging data, but none was
presented with MIR imaging data, which is a growing tool for the stellar census studies
in massive SFRs with the launch of \textit{Spitzer}. We demonstrate the ability of
high-resolution X-ray imaging to identify and understand the nature of the stellar
population in massive SFRs with a particular emphasis on the combination with NIR and
MIR imaging.

The outline of this paper is as follows. In \S~\ref{sect:s2}, we give a brief review of
RCW\,49 and Westerlund 2. In \S~\ref{sect:s3}, we present our X-ray and NIR observations
of this region. In \S~\ref{sect:s4}, we describe our X-ray and NIR data reduction and
the identification of X-ray sources with optical, NIR, and MIR sources using our deep
NIR data as well as archived data. In the data reduction, special care is taken in
source extraction, photometry, and cross-correlation in crowded regions. In total 468
X-ray sources are detected and 379 of them are identified by optical and infrared
sources.

In \S~\ref{sect:s5}, we discuss how the X-ray data are relevant to discriminate
intrinsic RCW\,49 members, identify new sources, and infer their nature. Seven
multi-faceted and loosely-related results are discussed in each subsection. In the first
subsection (\S~\ref{sect:s5-1}), we show that the central OB association is resolved in
the X-ray image and all the massive members in this region are detected in the X-ray. In
\S~\ref{sect:s5-2}, we argue that most of the X-ray sources with optical or infrared
counterparts are intrinsic to RCW\,49 by a comparison study with the X-ray population of
a control field in the Galactic Plane. In \S~\ref{sect:s5-3}, we obtain a loose
constraint on the distance to RCW\,49 using the mean X-ray luminosity of T Tauri stars
in the mass range of 2.0--2.7~$M_{\odot}$. In \S~\ref{sect:s5-4}, we evaluate the nature
of X-ray sources using X-ray and infrared photometric data. We employ the X-ray versus
NIR flux and luminosity plots in addition to the conventional NIR color-magnitude
diagram. We argue that the majority of X-ray sources are low-mass pre--main-sequence
sources and early-type main sequence stars. Some early-type stars are found to be harder
and more luminous in the X-ray than others with a similar NIR brightness. In
\S~\ref{sect:s5-5}, we study the spatial distribution of X-ray-identified cluster
members. Because of the low level of contamination of X-ray samples, we can identify
cluster members far from the central OB association. We find a possible excess
population at $\sim$4\arcmin\ and the cluster boundary at 6\arcmin--7\arcmin\ from the
cluster center. In \S~\ref{sect:s5-6}, we examine the fraction of RCW\,49 members
individually selected using X-ray emission and find that about half of the RCW\,49
sources brighter than \ksi$\sim$14~mag are identified by X-rays. The rate increases to
$\sim$90\% when combined with MIR excess emission, another indicator of cluster
membership complementary to the X-ray emission signature. In the last subsection
(\S~\ref{sect:s5-7}), we discuss the X-ray spectra of the five brightest X-ray
sources. We summarize these results in \S~\ref{sect:s6}.

\section{RCW\,49 and Westerlund 2}\label{sect:s2}
RCW\,49 was identified as an optically-visible \ion{H}{2} region by \citet{rogers60}. It
is also known as Kes\,10 \citep{kesteven68}, G\,284.3--00.3, and Gum\,29
\citep{gum55}. The visual extinction toward the region is \av\,$=$\,5--6~mag
\citep{carraro04,rauw07,ascenso07}. The stellar constituents in the region were studied
by imaging studies in the optical \citep{moffat75,moffat91,piatti98,carraro04,rauw07},
infrared \citep{churchwell04,whitney04,uzpen05,ascenso07}, and X-ray
\citep{goldwurm87,belloni94} bands. At least a dozen OB stars comprise the central OB
association called Westerlund 2 (Wd\,2; \citealt{westerlund60}). The association is also
known as NGC\,3247 \citep{dreyer88}, ESO\,127-18, and C\,1024-576.

\citet{moffat91} identified six O6--7 stars based on low-resolution optical
spectroscopy. \citet{rauw07} added six more O-type stars and gave a consistent
classification based on medium-resolution optical spectroscopy. These stars were
reclassified toward earlier spectral classes ranging between O3 and O6.5 \citep{rauw07}.

Three more early-type stars are found beyond the cluster core. One is a star with a
spectral type of O4\,V(f) or O5\,V(f) \citep{uzpen05,rauw07}, which we refer to as
MSP\,18 \citep{moffat91}. Another is a binary Wolf-Rayet star WR\,20a (WN6$+$WN6;
\citealt{vanderhucht01,rauw05}), which is the most massive binary in the Galaxy with a
well-determined mass \citep{rauw05}. Another Wolf-Rayet star WR\,20b
\citep{vanderhucht01} lies $\sim$3\farcm7 from Wd\,2 \citep{shara91}. These early-type
stars in and around Wd\,2 are probably the cause of exotic high energy phenomena
discovered in this region, such as extended hard X-ray emission \citep{townsley04} found
in the same \textit{Chandra} data set presented in this paper and TeV gamma-ray emission
\citep{aharonian07} found by the High Energy Stereoscopic System telescope.

A much larger number of lower-mass members are anticipated, many of which are at the
pre--main-sequence stage that can be identified either by infrared excess or X-ray
emission, arising respectively from circumstellar disks and envelopes and from enhanced
coronal magnetic activity. Some of the low-mass members were detected by mid-infrared
excess emission using the \textit{Spitzer} data \citep{whitney04}. In the X-ray band,
the two previous studies did not have sufficient spatial resolution and sensitivity to
distinguish individual stars. In these studies, respectively using \textit{Einstein} and
\textit{ROSAT}, only one \citep{goldwurm87} and seven \citep{belloni94} X-ray sources
were detected in the \textit{Chandra} studied field presented here. Most of these
sources are isolated early-type stars and unresolved stars in Wd\,2 (1E 1022.1--5730 in
\citealt{goldwurm87} and source number 24 in \citealt{belloni94}).

The ionized gas content at $\sim$10$^{4}$~K in RCW\,49 was studied by a radio continuum
imaging study at 0.843, 1.38, and 2.38~GHz \citep{whiteoak97}. Two shells appear in the
images, one around the complex of Wd\,2 and WR\,20a with a diameter of $\sim$7\farcm3
(radio ring A) and the other around WR\,20b with a diameter of $\sim$4\farcm1 (radio
ring B). Enhanced emission was found at the position where these two shells overlap. A
similar global structure was found in the dust content. The MIR image taken by the
Infrared Array Camera (IRAC) onboard \textit{Spitzer} traces the dust distribution via
polycyclic aromatic hydrocarbon emission at its three ([3.6], [5.8], and [8.0] $\mu$m)
of four bands \citep{churchwell04}. Moreover, the \textit{Spitzer} image reveals a
network of highly structured filaments, pillars, and shocks at arcsecond scales, some of
which are suggestive of instabilities.

Both gas and dust distributions are dominated by the massive stars in Wd\,2 and the two
Wolf-Rayet stars via their strong stellar winds and UV radiation that sculpt the
surrounding interstellar medium. These effects not only suppress further star formation,
but also trigger the second generation of star formation. From the spatial distribution
of MIR excess sources, \citet{whitney04} argued that star formation is taking place
preferentially in $\sim$4\arcmin\ around Wd\,2. The radius corresponds to the
outer boundary of the radio ring A where swept-up material is accumulated, suggesting
that these young sources are a consequence of induced star formation.

\medskip

The distance to RCW\,49 is highly controversial, ranging from 2 to 8~kpc in the
literature (\citealt{churchwell04} and references therein). The kinematic distance
determination gives a poor constraint because the region is in the tangential direction
of the Saggitarius-Carina arm. The debate on the distance continues in two most recent
papers; \citet{rauw07} argued for 8.0$\pm$1.4~kpc based on optical spectro-photometric
study of bright stars in Wd\,2, while \citet{ascenso07} proposed 2.8~kpc based on NIR
magnitudes and colors of RCW\,49 sources on the Henyey track. With the present X-ray
data, we can also loosely constrain the distance using the mass-stratified mean X-ray
luminosity (\S~\ref{sect:s5-3}). This gives an estimate of 2--5~kpc. Despite intensive
study, the distance to this cluster remains an open question. We adopt 4.2~kpc as used
by \citet{churchwell04}, which is roughly a geometric mean of the smallest and the
largest distance estimates, to avoid the worst case when the question is settled.

The association of Wolf-Rayet stars and an O4--O5\,V(f) star with Wd\,2 suggests that
the star cluster is no older than a few Myrs. No cluster member appears to be located
away from the zero-age--main-sequence isochrone curve in the optical color-magnitude
diagram, from which \citet{carraro04} estimated an age of $\lesssim$
2~Myr. \citet{ascenso07} similarly derived the cluster age of $\sim$2~Myr by fitting the
color-magnitude distribution of NIR sources with theoretical calculations.

\section{OBSERVATIONS}\label{sect:s3}
We carried out an X-ray observation of RCW\,49 using the Advanced CCD Imaging
Spectrometer (ACIS; \citealt{garmire03}) onboard the \textit{Chandra X-ray Observatory}
\citep{weisskopf02} from 2003 August 23 UT 18:20 to August 24 UT 4:54. Four imaging
array (ACIS-I) chips covered a 17\arcmin$\times$17\arcmin\ field centered at (R.\,A.,
decl.)$=$(10$^{\rm h}$24$^{\rm m}$00\fs5, --57\arcdeg45\arcmin18\arcsec) in the equinox
J2000.0 for a 36.7~ks exposure. ACIS-I covers the 0.5--8.0~keV energy band with a
spectral resolution of $\sim$150~eV at 6~keV and a point spread function (PSF) radius of
$\sim$0\farcs5 within $\sim$2\arcmin\ of the on-axis position, degrading to
$\sim$6\arcsec\ at a 10\arcmin\ off-axis angle. The data were taken with the very faint
telemetry mode and the timed exposure CCD operation with a frame time of 3.2~s.

We conducted near-infrared (NIR) observations on 2004 December 25 and 28 using the
Simultaneous three-color InfraRed Imager for Unbiased Surveys (SIRIUS;
\citealt{nagayama03}) mounted on the Cassegrain focus of the Infrared Survey Facility
(IRSF) 1.4~m telescope at the South African Astronomical Observatory. SIRIUS is a NIR
imager capable of obtaining simultaneous images in the \ji, \hi, and \ksi\ bands using
two dichroic mirrors. The instrument is equipped with three HAWAII arrays of
1024$\times$1024 pixels. The pixel scale of 0\farcs45 is an excellent match with the
on-axis spatial resolution of \textit{Chandra}. With ten-point dithering, we covered
8\farcm3$\times$8\farcm3 fields at two positions, one aimed at RCW\,49 (10$^{\rm
h}$24$^{\rm m}$01\fs9, --57\arcdeg45\arcmin31\arcsec) and the other at a control region
(10$^{\rm h}$27$^{\rm m}$13\fs5, --58\arcdeg01\arcmin26\arcsec) free from very bright
sources displaced by $\sim$30\arcmin\ from RCW\,49. Both fields have the same Galactic
latitude.

Each frame was exposed for 30~s and the total exposure time was $\sim$30~minutes equally
for the two regions. Additional short (5~s) exposure frames were also taken for the two
regions to extend the dynamic range of the data set to brighter sources. The observing
conditions were photometric with a seeing of $\lesssim$ 1\farcs0 on both nights.

\begin{figure*}[hbtp]
 \figurenum{1}
 \epsscale{0.5}
 \plotone{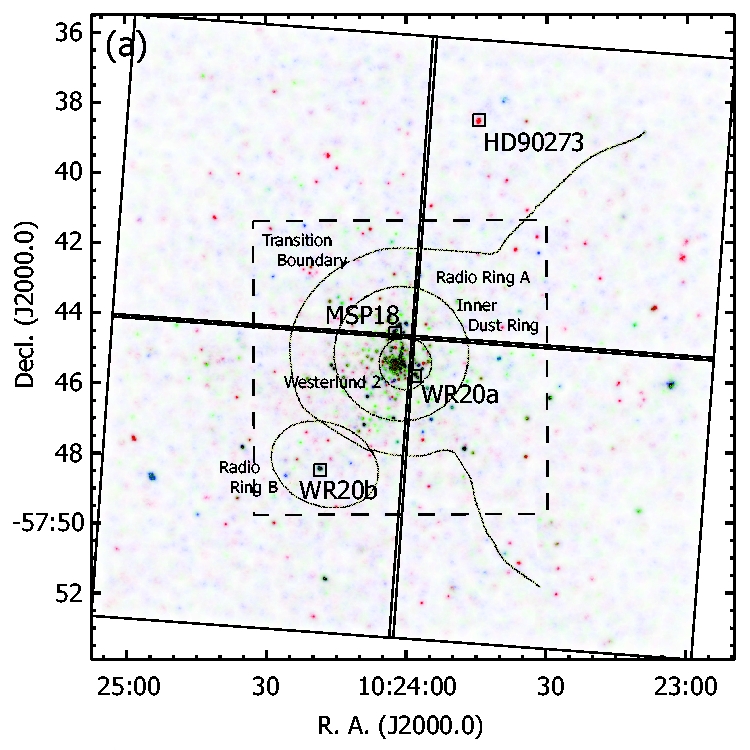}
 \plotone{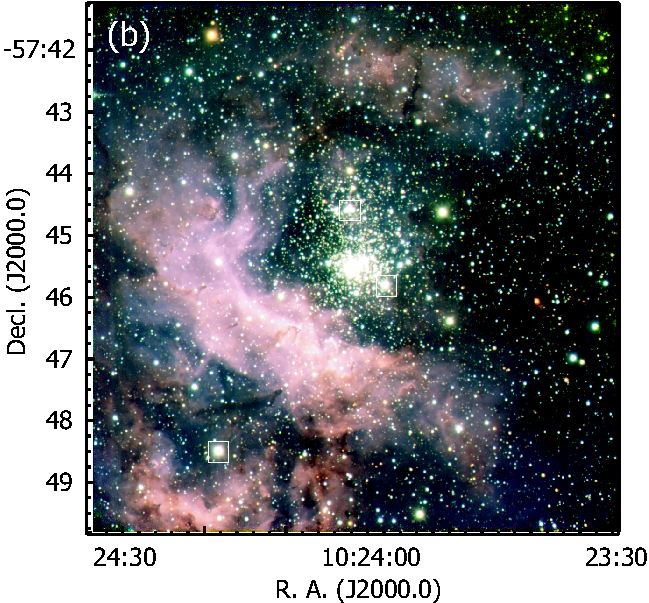}
 \caption{(a) Smoothed X-ray image. Red, green, and blue indicate the intensity in the
 0.5--1.7, 1.7--3.8, and 3.8--8.0~keV bands, respectively. The fields of view of the
 four ACIS chips and SIRIUS are shown by solid and dashed squares. Free-style lines
 taken from \citet{churchwell04} indicate the global features in the radio image
 \citep{whiteoak97}. (b) NIR image. Red, green, and blue indicate the intensity in the
 \ksi, \hi, and \ji\ bands, respectively. The positions of some landmark sources are
 given in small squares in both images and labeled in panel (a).}\label{fg:f1}
\end{figure*}

Figures~\ref{fg:f1} (a) and (b) respectively show the ACIS and SIRIUS images of the study
field. The SIRIUS image has a wider coverage by $\sim$2.9 times than the image
presented in \citet{ascenso07}.

\section{DATA REDUCTION AND ANALYSIS}\label{sect:s4}
\subsection{X-ray Data}
\subsubsection{Reduction \& Source Extraction}\label{sect:s4-1-1}
We reprocessed the raw data distributed by the \textit{Chandra} X-ray Center to obtain
an X-ray event list. A background rejection algorithm specific to data taken with the
very faint mode was applied. Events were further cleaned by removing cosmic-ray
afterglow and applying filters based on the event grades, status, and good-time
intervals \citep{townsley03}. Charge transfer inefficiency was corrected
\citep{townsley02} and the positions of all events were improved using a sub-pixel
repositioning technique \citep{tsunemi01}.

We detected sources using the wavdetect algorithm in the CIAO package\footnote{See
http://asc.harvard.edu/ciao/ for detail.} independently for the soft (0.5--2.0~keV),
hard (2.0--8.0~keV), and full (0.5--8.0~keV) band images. The software failed to detect
sources in crowded regions, where we manually added candidate sources with the aid of
smoothed images in different kernel sizes and PSF-deconvolved images obtained using a
maximum likelihood technique \citep{lucy74,townsley06}. In total, 556 X-ray source
candidates were identified including 60 manually added sources. We took a stance to add
sources manually in a liberal manner here, and we examined their validity in a rigorous
manner later.

For all the candidate sources, we used ACIS Extract\footnote{See the ACIS Extract User's
Guide at http://www.astro.psu.edu/xray/docs/TARA/ae\_users\_guide.html for detail. A
complete description of the procedure using the ACIS Extract package can be found in
\citet{getman05}.} version 3.107 for automated systematic source and background event
extractions. The source positions were refined by correlating the event distribution
with the PSF or by using the centroid of the events. Signals were then extracted from a
90\% encircled energy polygon of 1.5~keV X-rays to derive source counts in 0.5--8.0~keV
($C_{\rm{src}}$). When two extraction regions overlap, the regions were reduced. A
masked dataset was created by removing the events in a circle with a $\sim$99\%
encircled energy radius around all sources, from which background events were extracted
locally around each source to derive their background and net counts, $C_{\rm{bkg}}$ and
$C_{\rm{net}} = C_{\rm{src}} - C_{\rm{bkg}} \times A_{\rm{src}}/A_{\rm{bkg}}$, where
$A_{\rm{src}}$ and $A_{\rm{bkg}}$ are the integrals of the masked exposure map values
within the source and background extraction regions.

Each source candidate was tested for validity using the photometric significance (PS)
and probability of no source ($P_{\rm{B}}$) statistics extracted by ACIS Extract. PS is
the commonly used photometric signal-to-noise ratio given by
\begin{equation}
 \mathrm{PS} = \frac{C_{\mathrm{net}}}{\Delta C_{\mathrm{net}}},
\end{equation}
where the approximation by \citet{gehrels86} was employed to calculate the uncertainty
$\Delta C_{\rm{net}}$. $P_{\rm{B}}$ is the probability for the null hypothesis that all
detected counts or more are explained by fluctuations of the background and is defined
by
\begin{equation}
 P_{\mathrm{B}} = 1 - \sum\limits_{n=0}^{C_{\mathrm{src}}-1} P(C_{\mathrm{bkg}} \times A_{\mathrm{src}}/A_{\mathrm{bkg}}, n),
\end{equation}
where $P(\lambda,n)$ indicates a Poisson probability distribution function of a mean
$\lambda$ to have $n$ counts.

We recognize 468 source candidates with $\mathrm{PS}$~$\ge$~1.0 and
$P_{\rm{B}}$~$\le$~1.0$\times$10$^{-2}$ to be valid X-ray sources
(Table~\ref{tb:t1}). None of them is bright enough to cause pile-up. Almost all the
sources are newly detected here.

\subsubsection{Photometry \& Time Variability}
For all valid sources, ACIS Extract was used for recalculating their source and
background counts, producing instrumental responses, and constructing spectra and light
curves for systematic photometry and spectroscopy analyses. The photometry results are
compiled in Table~\ref{tb:t1}. Columns (1) and (2) are the sequential number and the
source name. Columns (3)--(6) give position information (R.\,A. and decl.\, in the
equinox J2000.0, their uncertainty, and the off-axis angle). Columns (7)--(9) show
counts in 0.5--8.0~keV; $C_{\rm{net}}$, $\Delta C_{\rm{net}}$, and
$C^{\prime}_{\rm{bkg}} = C_{\rm{bkg}} \times A_{\rm{src}}/A_{\rm{bkg}}$, whereas column
(10) shows the net counts in 2.0--8.0~keV ($C_{\rm{net,hard}}$). Column (11) indicates
the PSF fraction of source extraction, in which smaller numbers than the default value
($\sim$0.9) indicate that the extraction region was reduced to avoid
overlapping. Columns (12) and (13) are for PS and $P_{\rm{B}}$. Columns (14) and (15)
are sets of flags to indicate extraction anomalies and source variability. Column (16)
gives the effective exposure time, the time needed to accumulate the observed counts if
the source were at the aim point. Column (17) lists the median energy of the extracted
X-ray events corrected for the background. Median energy is proposed to be a better
quantity to characterize X-ray spectral hardness and to give a reliable estimate of
extinction than the conventionally-used hardness ratio
\citep{hong04,feigelson05}. Column (18) is the photometric flux discussed in
\S~\ref{sect:s4-1-3}.

X-ray variability was examined for 377 bright sources off the chip gaps using a
Kolmogorov-Smirnov test. The null hypothesis probability ($P_{\rm{KS}}$) that the flux
is constant was tested and sources were placed into three classes; (a) no evidence for
variability ($P_{\rm{KS}} > 5 \times 10^{-2}$), (b) possibly variable ($5 \times 10^{-3}
< P_{\rm{KS}} \le 5 \times 10^{-2}$), and (c) definitely variable ($P_{\rm{KS}} \le 5
\times 10^{-3}$). Seven sources are in the ``definitely variable'' class, showing
flare-like light curves.

\subsubsection{Spectroscopy \& Flux Estimates of Faint Sources}\label{sect:s4-1-3}
Table~\ref{tb:t2} gives the results from the X-ray spectral fits obtained with ACIS
Extract, where 230 spectra with PS $\ge$ 2 were fitted by an optically-thin thermal
plasma (APEC) model \citep{smith01} with interstellar absorption. The best-fit values of
plasma temperature (\kt), the volume emission measure (\emm), and the extinction column
density (\nh) were derived. The photometric and spectroscopic tables (Tables~\ref{tb:t1}
and \ref{tb:t2}) are in a format similar to those found in associated papers
\citep{getman06,townsley06,getman07a,jwang07a,broos07,jwang07b}.

\begin{figure}[hbtp]
 \figurenum{2}
 \epsscale{1.0}
 \plotone{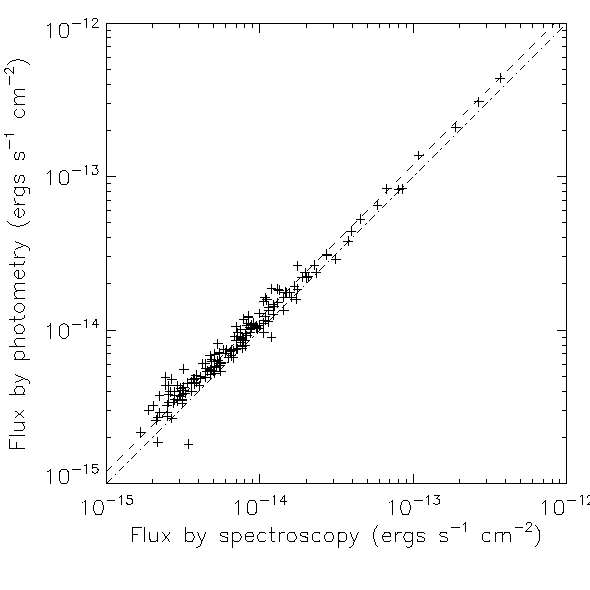}
 \caption{Consistency between the two X-ray flux estimates in the 0.5--8.0~keV band. The
 values derived from spectral fits (abscissa) are plotted against those from broad-band
 photometry (ordinate) for 173 sources with successful thermal fits in
 Table~\ref{tb:t2}. The dashed-and-dotted line indicates $y=x$, while the dashed line
 indicates the best-fit model of $y=1.20x$.}\label{fg:f2}
\end{figure}

The X-ray flux can be derived by the spectral fits for most of the sources in
Table~\ref{tb:t2}. For the remaining sources, which are mostly faint, we estimated their
X-ray flux using broad-band photometry: the mean energy times the net count rate divided
by the mean effective area \citep{tsujimoto05}. The consistency between the two flux
estimates was checked for the brightest 173 sources (Fig.~\ref{fg:f2}), for which both
values are available. They are overall in good agreement with a systematic offset
(20\%). We hereafter use the spectroscopic flux for the 173 bright sources and the
photometric flux corrected for the offset for the remaining sources.

\subsection{NIR Data}
All SIRIUS frames were reduced following standard procedures using IRAF, including
dark-current correction, sky subtraction, and flat-fielding. Dithered images were
registered to construct combined images in the three bands (\ji, \hi, and \ksi). Sources
were extracted and their photometry was measured using the daofind and daophot tasks in
IRAF.

In order to cope with source crowding in the SIRIUS images, we used the following
procedure \citep{stetson87}. First, we constructed the average PSF from dozens of bright
sources across each image. The profile of each extracted source was fitted with the
average PSF and the source was masked from the image if the fit was successful. The
procedure was repeated until most sources were masked. We then reconstructed the PSF
from the masked image, which we consider is freer from contamination than the original
one. By repeating the procedure, we separated closely spaced sources and deblended their
overlapping signals.

We extracted 10,540 and 9,768 sources in RCW\,49 and the control fields, respectively,
at levels larger than 3 $\sigma$ of the sky background noise in all three bands. We
derived their positions and magnitudes in the \ji, \hi, and \ksi\ bands in the
longer-exposure images. We replaced magnitudes with those derived from the
shorter-exposure images when they were saturated at levels brighter than 11.0 (\ji),
11.5 (\hi), and 12.0 (\ksi) mag. The 3 $\sigma$ detection limits of the longer-exposure
images in the RCW\,49 field are $\sim$20.5 (\ji), $\sim$19.8 (\hi), and $\sim$18.4
(\ksi) mag, whereas the 10 $\sigma$ limits are $\sim$19.0 (\ji), $\sim$17.8 (\hi), and
$\sim$16.8 (\ksi) mag. These limits were $\sim$0.5~mag shallower than those measured in
the control field because of the larger source confusion. All the detected sources are
used for the X-ray counterpart search, while only those with magnitude uncertainty less
than 0.1~mag are used for photometric analysis.

\subsection{Optical and Infrared Identifications of X-ray Sources}
We identified the X-ray sources in the optical, NIR, and MIR bands using the Naval
Observatory Merged Astrometric Dataset (NOMAD)\footnote{See
http://www.nofs.navy.mil/nomad/ for details.}, the Two-Micron All-Sky Survey (2MASS;
\citealt{skrutskie97}), SIRIUS, and GLIMPSE. GLIMPSE is a Legacy Program of the
\textit{Spitzer Space Telescope} to survey a $\sim$220 deg$^{2}$ region of the Galactic
Plane at four MIR bands ([3.6], [4.5], [5.8], and [8.0] $\mu$m) with a
1\farcs6--1\farcs9 resolution using IRAC \citep{fazio04}. A 1\fdg7$\times$0\fdg7 region
encompassing RCW\,49 was mapped with ten 1.2~s exposures in the initial phase of the
program \citep{churchwell04}.

\begin{figure}[hbtp]
 \figurenum{3}
 \epsscale{1.0}
 \plotone{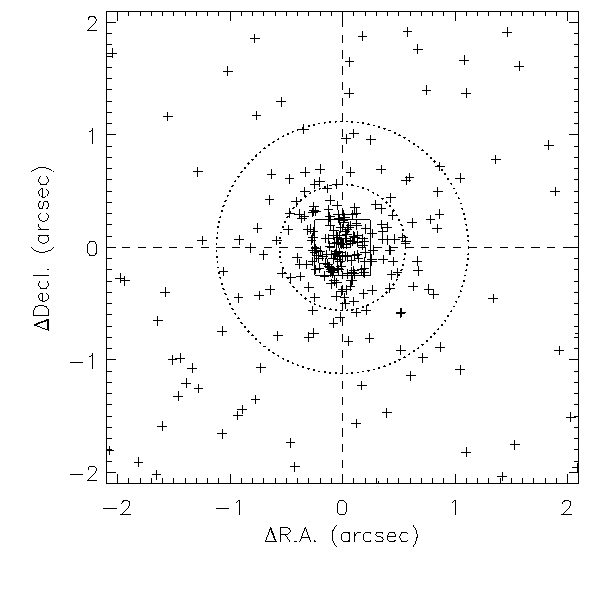}
 \caption{Astrometric consistency between ACIS and 2MASS after correcting for a
 systematic offset. The differences of R.\,A. and decl. between the ACIS and 2MASS
 positions ($\Delta$R.\,A. and $\Delta$Decl.) are plotted for all the closest pairs. The
 dotted circles are drawn at 1 and 2 $\sigma$ radius, whereas the solid square
 represents the ACIS pixel size of 0\farcs492$\times$0\farcs492.}\label{fg:f3}
\end{figure}

First, the ACIS image was shifted to match the absolute astrometry of 2MASS using the
closest ACIS--2MASS pairs. The residual displacements of the pairs in the R.\,A. and
decl. directions are plotted in Figure~\ref{fg:f3}, in which the rms (1~$\sigma$) of the
displacements is $\sim$0\farcs56 (Fig.~\ref{fg:f3}; the inner dotted circle). We
recognized 216 ACIS--2MASS pairs within 2 $\sigma$ displacement (Fig.~\ref{fg:f3}; the
outer dotted circle) to be physical counterparts. We similarly found NOMAD, SIRIUS, and
IRAC counterparts within 2 $\sigma$ of ACIS sources after a correction of the systematic
offset, where 1 $\sigma$ is $\sim$0\farcs63, $\sim$0\farcs38, and $\sim$0\farcs59,
respectively. Consequently, 230, 299, and 156 ACIS sources were found to have NOMAD,
SIRIUS, and IRAC counterparts. Both the numbers of false positives (unrelated pairs
identified as counterparts) and false negatives (physical pairs unidentified as
counterparts) are estimated to be $<$10.

The resultant identifications and the optical (\ri), NIR (\ji, \hi, and \ksi), and MIR
([3.8], [4.5], [5.8], and [8.0]) photometry are summarized in
Table~\ref{tb:t3}. The \ri-band magnitude is from NOMAD, which is a compilation work of
several catalogues. NIR magnitudes are either from 2MASS or SIRIUS, which is indicated
by the NIR flag column.

In total, 379 of the 468 ACIS sources have either NOMAD, 2MASS, SIRIUS, or IRAC
counterparts. Except for massive stars in and around Wd\,2, almost all the X-ray sources
are poorly studied in other wavelengths. Spectroscopic classifications are available for
five sources outside of Wd\,2 \citep{moffat91,rauw07}: HD\,90273 (O\,7), HD\,302752 (A),
MSP\,91 (G0\,V--III), MSP\,158 (G0\,III), and MSP\,218 (A0\,III), among which only HD\,90273
was detected in X-rays. The identifications with sources in earlier work are also given
in Table~\ref{tb:t3}.

\section{DISCUSSION}\label{sect:s5}
\subsection{Early-type stars in Westerlund 2 and Beyond}\label{sect:s5-1}
\begin{figure*}[hbtp]
 \figurenum{4}
 \epsscale{0.5}
 \plotone{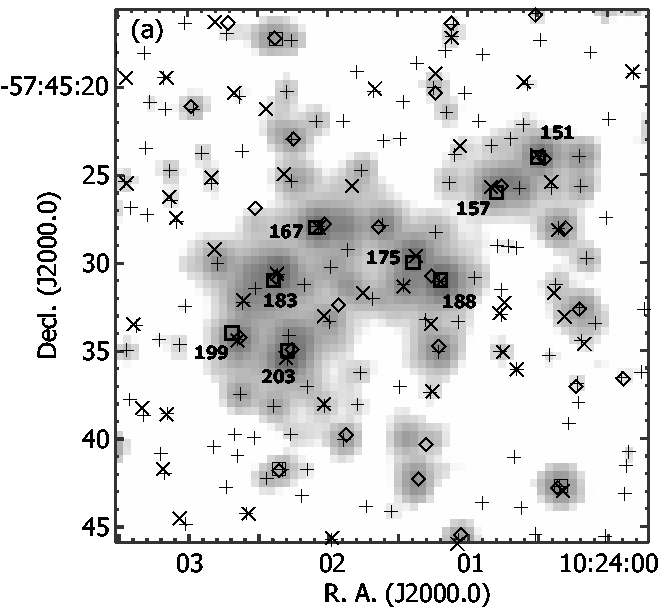}
 \plotone{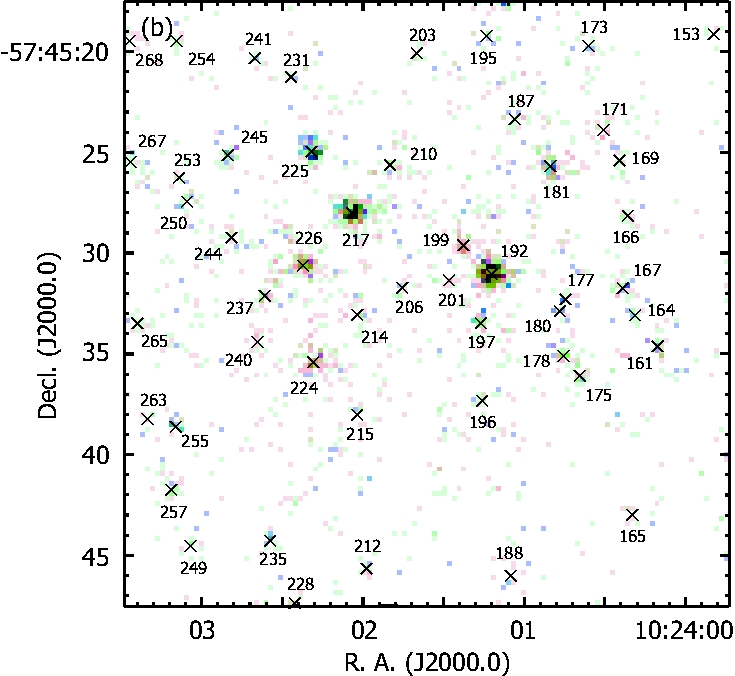}
 \caption{Close-up views of Wd\,2. (a) SIRIUS \ksi-band image. The positions of ACIS
 (crosses), SIRIUS (pluses), IRAC (diamonds), and 2MASS (squares) sources are shown.
 O-type stars in \citet{rauw07} are also shown by thick squares with their names (the
 prefix ``MSP'' is omitted). (b) \textit{Chandra} image. The color coding is the same
 as in Figure~\ref{fg:f1} (a). The positions of Chandra sources are shown by
 crosses with their sequential numbers (Table~\ref{tb:t1}).}\label{fg:f4}
\end{figure*}

Figure~\ref{fg:f4} shows a close-up view of Wd\,2 in the SIRIUS and ACIS images. The
\textit{Chandra} observation of RCW\,49 resolved X-ray sources in the central OB
association. A dozen early-type stars from O7 to O3 are present in the cluster core ;
MSP\,151, 157, 167, 171, 175, 182, 183, 188, 199, 203, and 263 \citep{rauw07}. In
addition, three early-type stars, an O4\,V(f) or O5\,V(f) star (MSP\,18;
\citealt{uzpen05,rauw07}) and two Wolf-Rayet stars (WR\,20a $=$ MSP\,240 and WR\,20b),
lie outside the cluster core.

We detected X-ray emission from all of these early-type sources. The identifications are
given in Table~\ref{tb:t3}. The results illustrate clearly that high-resolution X-ray
imaging observations, even with a short exposure of 30--40~ks, are powerful enough to
resolve and identify the crowded massive members in OB associations at a few kiloparsecs
away, as was found by \citet{jwang07a} in NGC\,6357, \citet{broos07} in M\,17, and
similar studies.

\subsection{Galactic \& Extra-Galactic Contaminants}\label{sect:s5-2}
The population of X-ray sources detected in ACIS images of Galactic massive SFRs is a
superposition of the young cluster members, field stars in the Galactic Plane, and
extragalactic background. In order to examine how many X-ray sources toward RCW\,49 are
intrinsic to the cluster, we compare the RCW\,49 population with that in a control field
of the Galactic Plane \citep{ebisawa05}.

\begin{figure}[hbtp]
 \figurenum{5}
 \epsscale{1.0}
 \plotone{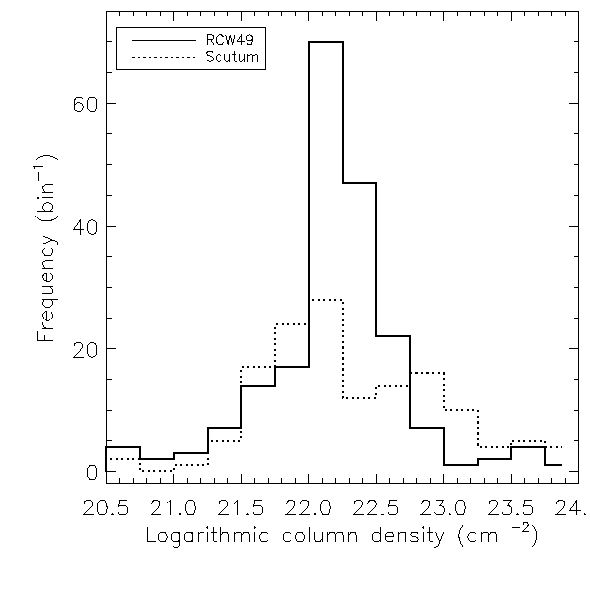}
 \caption{Histograms of the \nh\ values of X-ray sources derived by spectral fits
 (Table~\ref{tb:t2}). The solid and dashed histograms are for the RCW\,49 and the Scutum
 regions, respectively.}\label{fg:f5}
\end{figure}

The control field, the ``Scutum region'' at ($l$, $b$) $\sim$ (28\fdg5, 0\fdg0), is an
ideal place for our comparison study. First, the region was covered by X-ray and NIR
imaging studies with a sensitivity similar to RCW\,49. Second, it does not harbor a
young stellar cluster and the X-ray population there is a typical Galactic Plane
population superposed on obscured extragalactic contaminants. Third, the extinction in
the two regions is similar. The typical \nh\ in RCW\,49 is derived to be
10$^{22}$--10$^{22.5}$~cm$^{-2}$ from Figure~\ref{fg:f5}, which is consistent with the
\textit{ROSAT} \citep{belloni94} and NIR extinction \citep{ascenso07} measurements. The
Scutum region shows a bimodal \nh\ distribution unlike RCW\,49 (Fig.~\ref{fg:f5}), but
their median values are the same (10$^{22.2}$~cm$^{-2}$). In any case, the extinction
correction is not very important in the hard X-ray band, where the opacity is less than
one.

\begin{figure}[hbtp]
 \figurenum{6}
 \epsscale{1.0}
 \plotone{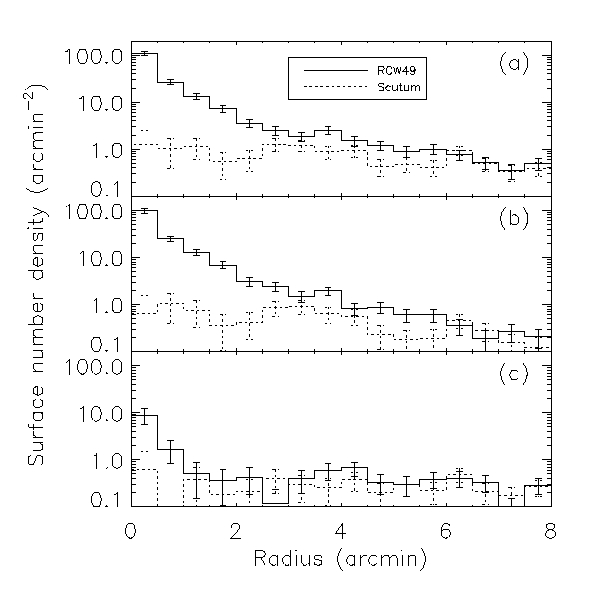}
 \caption{Radial SND profile of (a) all X-ray sources, (b) sources with optical, NIR, or
 MIR counterparts, and (c) those without counterparts. Solid and dashed histograms
 indicate the data from RCW\,49 and the Scutum region. The center of the profiles is the
 mean position of all X-ray sources in RCW\,49 and the same detector position for the
 Scutum region. The uncertainty is estimated by $\sqrt{N}$, where $N$ is the number of
 sources in each bin.}\label{fg:f6}
\end{figure}

We use the radial profile of X-ray surface number density (SND) in RCW\,49 to argue that
a large fraction of our X-ray sources consists of the intrinsic RCW\,49
population. First, the radial SND profile of all X-ray sources (Fig.~\ref{fg:f6}a) shows
a centrally peaked shape. The peak at R.\,A.$=$10:24:01.6 and decl.$=$--57:45:31, which
was derived as the mean position of all X-ray sources, is consistent with the position
of Wd\,2. Second, the SND of RCW\,49 has an overpopulation compared to the Scutum region
(Fig.~\ref{fg:f6}a; \textit{dotted histogram}) out to $\sim$6\arcmin--7\arcmin. Within
5\arcmin\ of the cluster center where we have SIRIUS coverage, the number of X-ray
sources is 374 while the projected number of the Scutum region is $\sim$69, indicating
that $\sim$82\% of the X-ray sources are intrinsic to RCW\,49.

Our claim that the X-ray population suffers little contamination by non--cluster-members
is more evident for those with stellar (optical, NIR, and MIR) identifications
(Fig.~\ref{fg:f6}b). We have 330 identified X-ray sources within 5\arcmin\ of the Wd\,2
center, while the projected number in the Scutum region is $\sim$47, which accounts only
for $\sim$14\% of the RCW\,49 count.

\begin{figure}[hbtp]
 \figurenum{7}
 \epsscale{1.0}
 \plotone{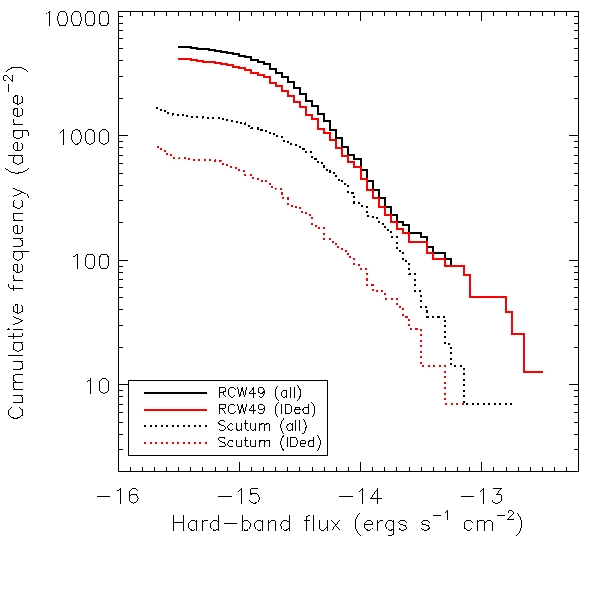}
 \caption{$\log{N}-\log{S}$ plots of RCW\,49 (solid) and the Scutum region
 (broken). Hard-band absorption-uncorrected flux was used. Histograms are constructed
 for all sources (black) and those with stellar counterparts (red),
 respectively.}\label{fg:f7}
\end{figure}

This overpopulation holds in the entire X-ray flux range of the \textit{Chandra}
study. In the comparison of the hard-band $\log{N}-\log{S}$ relations
(Fig.~\ref{fg:f7}), we find the total and identified X-ray populations of the RCW\,49
region (\textit{thick histograms}) are well above those of the Scutum region
(\textit{broken histograms}).

\subsection{Distance Constraint from X-ray Data}\label{sect:s5-3}
The distance to RCW\,49 is controversial despite intensive studies across many
wavelengths (\S~\ref{sect:s2}). In the present study, we obtained a rich X-ray sample
and learned that most of the identified X-ray sources are intrinsic to RCW\,49. It is
therefore worthwhile to attempt to constrain the distance with the new X-ray data set.

We use a method proposed in \citet{getman07b} to constrain the distance based on X-ray
data. In this method, a constraint comes from the dependence of X-ray luminosity on the
mass of late-type pre--main-sequence sources. The relation between these two stellar
quantities is established consistently in several young star clusters, including Cepheus
B \citep{getman06}, the Orion Nebula Cluster (\citealp[Table~6]{getman06};
\citealt{preibisch05}), the Taurus Molecular Cloud \citep {guedel07}, and CG\,12
\citep{getman07b}. Note that the X-ray luminosity versus mass relation is not degenerate
with respect to distance because pre--main-sequence X-ray luminosity and mass depend
differently on the assumed distance.

First, we estimate the stellar mass of X-ray sources, for which NIR photometry is
available (Table~\ref{tb:t3}). We construct a \ji/(\ji--\hi) color-magnitude diagram
assuming various distances from 0.5 to 8~kpc and derive the mass by comparing the
dereddened \ji-band magnitudes to the \citet{siess00} model. Then, we extract the X-ray
sources with an estimated mass of 2.0--2.7~$M_{\odot}$ and derive the median value of
their logarithmic hard-band absorption-corrected X-ray luminosity ($\log L_{\rm{h,c}}$
in Table~\ref{tb:t2}) in order to minimize the extinction effect. The mass range is
limited by the degeneracy in the mass estimate above $\sim$2.7~$M_{\odot}$ and by our
X-ray sensitivity below $\sim$2.0~$M_{\odot}$.

\begin{figure}[hbtp]
 \figurenum{8}
 \epsscale{1.0}
 \plotone{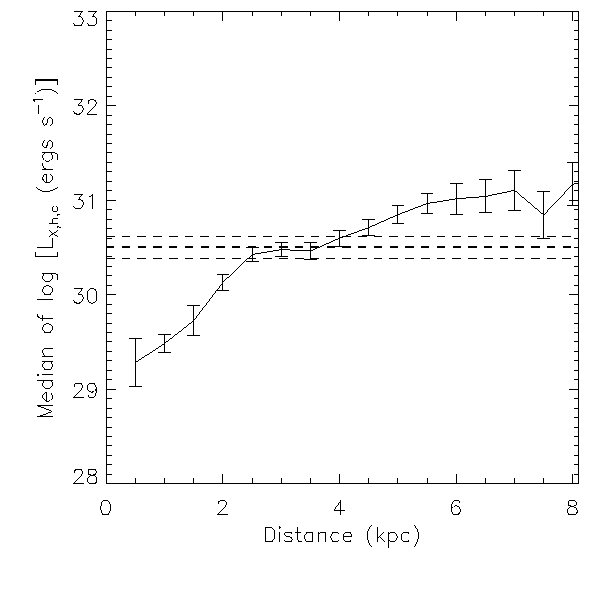}
 \caption{Median X-ray luminosity versus assumed distance of RCW\,49 sources. The
 logarithmic median value of hard-band absorption-corrected X-ray luminosities in the
 mass ranges of 2.0--2.7~$M_{\odot}$ is plotted at each assumed distance. The error bars
 show the median absolute dispersion (MAD) divided by the number of samples (7--46
 sources) at each bin, which represent the uncertainty of median \lx\ determination. The
 dotted and dashed lines indicates the median \lx\ and its MAD divided by the number of
 samples from the Orion Nebula Cluster in the same mass range, respectively
 \citep{getman06}.}\label{fg:f8}
\end{figure}

Figure~\ref{fg:f8} shows the median X-ray luminosity at each assumed distance, in which
the estimated mean X-ray luminosity increases monotonically as the distance
increases. We compare the median X-ray luminosity in the same mass range of the Orion
Nebula Cluster \citep{getman06}, for which the distance is reliably measured. We found
that the distance is constrained to be 2--5~kpc, which is consistent with the claim by
\citep{ascenso07} but not with \citep{rauw07}.

\subsection{Nature of X-ray Cluster Population}\label{sect:s5-4}
\begin{figure*}[hbtp]
 \figurenum{9}
 \epsscale{0.5}
 \plotone{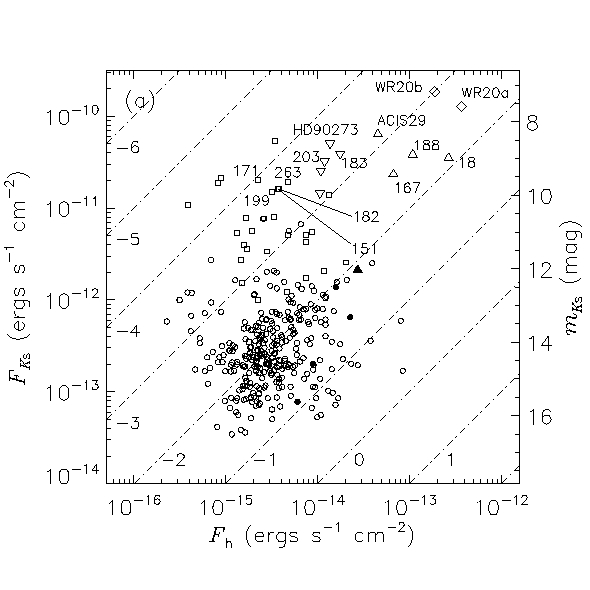}
 \plotone{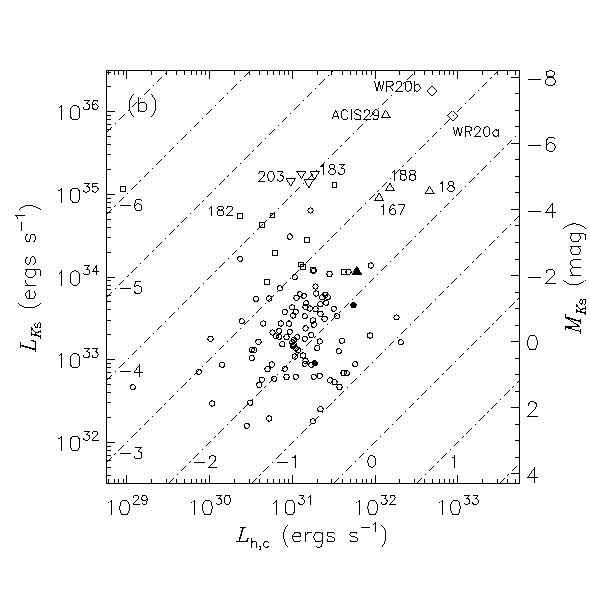}
 \caption{Scatter plots of X-ray and NIR brightness of NIR-identified X-ray sources. (a)
 Hard-band X-ray flux ($F_{\rm{h}}$) versus \ksi-band flux ($F_{K\rm{s}}$) or magnitude
 ($m_{K\rm{s}}$) for 339 X-ray sources with \ksi-band counterparts. (b)
 Extinction-corrected hard-band X-ray luminosity ($L_{\rm{h,c}}$) versus \ksi-band
 luminosity ($L_{K\rm{s}}$) or absolute magnitude ($M_{K\rm{s}}$) for 124 X-ray sources
 with \nh\ estimates in Table~\ref{tb:t2}. Sources are NIR-loud toward the top left and
 X-ray-loud toward the bottom right of the figures. Wolf-Rayet stars are shown by large
 diamonds. Sources brighter than B2\,V in NIR are shown by upward triangles for bright
 (more than 50 counts) and hard (best-fit temperature larger than 2~keV), downward
 triangles for bright and soft, and squares for faint (less than 50 counts)
 sources. Sources fainter than B2\,V in NIR are indicated by circles. Those with X-ray
 variability are marked with filled symbols. Spectroscopically-identified early-type
 stars are labeled. The prefix ``MSP'' is omitted for \citet{moffat91} sources. ACIS\,29
 indicates the source number 29 in Tables~\ref{tb:t1}, \ref{tb:t2}, and \ref{tb:t3}. The
 dashed-and-dotted lines indicate the iso-$x/y$ lines from 10$^{-7}$ to 10$^{2}$ (the
 logarithmic values are given beside each line).} \label{fg:f9}
\end{figure*}

Sources of different natures have different X-ray and NIR brightness and colors. In
massive SFRs, where the mass range of stellar constituents is wider than low-mass SFRs,
the brightness is of prime importance for estimating the nature. We employ X-ray versus
NIR brightness plots (Figs.~\ref{fg:f9}a and \ref{fg:f9}b) to estimate the nature of
X-ray cluster members of RCW\,49, aided by the conventional NIR color-magnitude diagram
(Fig.~\ref{fg:f11}a).

Figure~\ref{fg:f9} (a) shows the observed \ksi-band magnitude ($F_{K\rm{s}}$) versus
hard-band X-ray flux ($F_{\rm{h}}$) of all the X-ray sources with \ksi-band
identification. Figure~\ref{fg:f9} (b) is restricted to a sample with X-ray spectral
fits, showing the extinction-corrected absolute \ksi-band luminosity ($L_{K\rm{s}}$)
versus hard-band X-ray luminosity ($L_{\rm{h,c}}$). The constant flux ratio
lines between the \ksi-band and the hard X-ray band (\textit{dotted-and-dashed lines} in
Figure~\ref{fg:f9}a and b) are a proxy for the canonical X-ray versus bolometric
luminosity, which is independent from our distance estimate.

It is easily recognized that the sources form several groups in both plots. The most
populous group centered at ($F_{\rm{h}}$, $F_{K\rm{s}}$) $\sim$ ($10^{-14.8}$,
$10^{-12.7}$)~ergs~s$^{-1}$~cm$^{-2}$ in Figure~\ref{fg:f9} (a) and ($L_{\rm{h.c}}$,
$L_{K\rm{s}}$) $\sim$ ($10^{31}$, $10^{33.5}$)~ergs~s$^{-1}$ in Figure~\ref{fg:f9} (b)
are 1.0--3.0~$M_{\odot}$ low-mass pre--main-sequence sources in RCW\,49 for the
following reasons. Their \ksi-band brightness (13--16~mag) is consistent with the NIR
brightness of pre--main-sequence sources in this mass range. Their hard-band X-ray
luminosities 10$^{30}$--10$^{31.5}$~ergs~s$^{-1}$ and flaring are typical of this class
of stars in near-by regions (e.g., \citealt{wolk06,getman06}). The ratio of X-ray to NIR
brightness of 10$^{-1}$--10$^{-3}$ is consistent with the fractional X-ray to bolometric
luminosity of 10$^{-3}$--10$^{-5}$ common among these sources. The \ksi-band bolometric
corrections for A0\,V, F0\,V, and G0\,V stars are --3.75, --2.85, and --2.25~mag
\citep{tokunaga00,drilling00}.

A number of sources deviate in the upward and rightward directions of the primary group
in Figure~\ref{fg:f9}. Among them, two Wolf-Rayet stars occupy the top right end of all
sources with ($F_{\rm{h}}$, $F_{K\rm{s}}$) $\sim$ ($10^{-12.8}$,
$10^{-10.2}$)~ergs~s$^{-1}$~cm$^{-2}$ in Figure~\ref{fg:f9} (a) and ($L_{\rm{h,c}}$,
$L_{K\rm{s}}$) $\sim$ ($10^{33}$, $10^{36}$)~ergs~s$^{-1}$ in Figure~\ref{fg:f9} (b).

\begin{figure*}[hbtp]
 \figurenum{10}
 \epsscale{0.5}
 \plotone{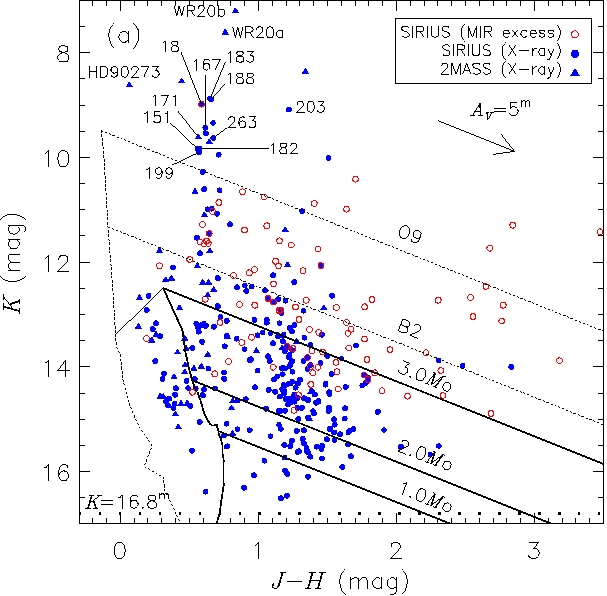}
 \plotone{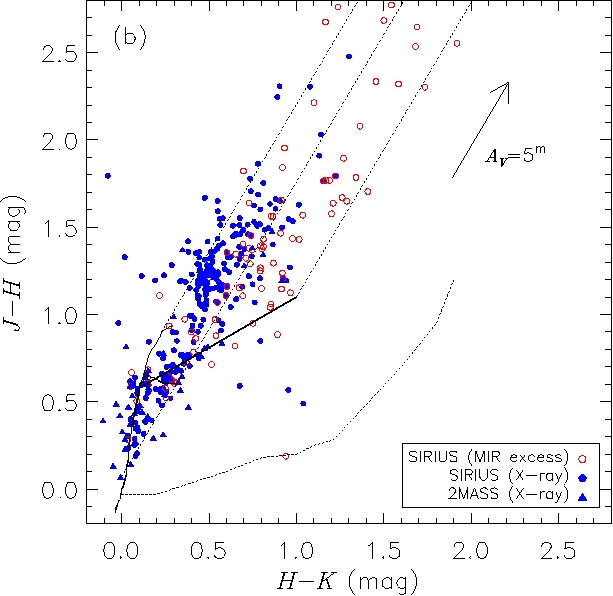}
 \caption{NIR diagrams of RCW\,49. SIRIUS sources with X-ray and MIR excess emission are
 respectively shown by blue filled and red open circles. X-ray sources with 2MASS but
 without SIRIUS photometry are plotted by blue filled triangles. 2MASS and SIRIUS colors
 are transformed into the California Institute of Technology color system to compare
 with the theoretical calculations using the formulae given respectively by
 \citet{carpenter01} and \citet{nakajima05}. (a) Color-magnitude diagram. A 2~Myr
 isochrone curve of pre--main-sequence stars (thick solid curve) is from \citet[$M <
 1.4~M_{\odot}$]{baraffe98} and \citet[$3.0~M_{\odot} > M >
 1.4~M_{\odot}$]{siess00}. The reddening lines of 1, 2, and 3~$M_{\odot}$ sources are
 shown with solid lines. A color-magnitude curve of main sequence stars (thick dashed
 curve) is from \citet{tokunaga00,drilling00}. The reddening lines of O9\,V and B2\,V
 stars are shown with dashed lines. The SIRIUS \ki-band magnitude limit is indicated by
 the dotted line. Spectroscopically-identified early-type stars are labeled. The prefix
 ``MSP'' is omitted for \citet{moffat91} sources. (b) Color-color diagram. The intrinsic
 colors of dwarfs and giants (thick curves) are from \citet{tokunaga00}, the classical T
 Tauri star locus (thick line) is from \citet{meyer97}, and the intrinsic colors of
 Herbig Ae Be stars (dotted curve) are from \citet{lada92}.}\label{fg:f10}
\end{figure*}

The sources in between the late-type pre--main-sequence and Wolf-Rayet star groups, both
in X-ray and NIR brightness, are early-type main-sequence stars. They have a large
dispersion in the horizontal direction in Figure~\ref{fg:f9}, indicating that sources
with a similar NIR brightness can differ significantly in hard X-ray brightness. We
first extract early-type star candidates based on the NIR color-magnitude diagram
(Fig.~\ref{fg:f10}a). X-ray sources with dereddened \ki-band magnitude brighter than a
B2\,V star are considered to be early-type candidates and are labeled ``ET'' in
Table~\ref{tb:t3}. A total of 43 candidate stars include 13 spectroscopically-identified
O-type stars from preceding studies \citep{moffat91,rauw07}. These early-type candidates
are plotted by different symbols based on their X-ray spectral hardness.

The upward and downward triangle sources, which respectively indicate hard and soft
X-ray emission by X-ray spectroscopy, occupy different regions naturally reflecting the
fact that harder O stars are brighter in the hard X-ray band. The two groups appear to
form two separate clustering in Figure~\ref{fg:f9} and have different X-ray to NIR
brightness ratio. These suggests that they have different X-ray production
mechanisms. Eight spectroscopically-identified sources (HD\,90273, MSP\,151, 171, 182,
183, 199, 203, and 263) are in the soft O-star group with $L_{\rm{h,c}}$/$L_{K\rm{s}}
\sim 10^{-4}$, while three (MSP\,18, 167, and 188) are in the hard O-star group with
$L_{\rm{h,c}}$/$L_{K\rm{s}} = 10^{-2} - 10^{-3}$. Source number 29 (ACIS 29) may belong
either to the hard O-star or Wolf-Rayet star group, but the nature of this source is
unknown without any spectroscopic studies. We note that the source has infrared excess
emission (Table~\ref{tb:t3}) and is an off-cluster source at $\sim$4\farcm5 away from
the center, which is similar to WR\,20a and MSP\,18.

The soft O stars share common characteristics with traditional X-ray-emitting O stars
\citep{berghoefer97} for having soft and constant X-ray emission due to internal shocks
in their stellar winds \citep{lucy80}. The hard O stars are among the growing class of
new-type O stars with hot thermal emission (\S~\ref{sect:s5-7-2}). The three hard O
stars (MSP\,18, 167, and 188) are bright enough for detailed X-ray spectroscopy, and we
discuss them in \S~\ref{sect:s5-7-2}.

\subsection{Spatial Distribution of X-ray Cluster Members}\label{sect:s5-5}
We can individually identify cluster members based on their X-ray identification with
a low level of contamination. Unlike optical and infrared imaging, the X-ray detections
are not affected by non-uniform and intense diffuse emission. These two features of
X-ray imaging studies are particularly important outside of the cluster core, where the
optical and NIR SNDs of cluster members fall to a comparable level to those of
irrelevant sources. Using the X-ray SND profiles, we obtain a sensible estimate of the
spatial scale and distribution of cluster members away from the cluster core.

In the X-ray SND profile (Fig.~\ref{fg:f6}a), we see two interesting features. One is
that the profile shows an overall decline, reaching the background level at
6\arcmin--7\arcmin. The X-ray declining profile is smoother and more extended than the
NIR profile constructed from 2MASS \citep[Fig.~6]{ascenso07}. We obtained a larger
estimate of the cluster size by twice than the NIR estimate of 4\farcm1\
\citep{ascenso07}. Our estimated radius is $\sim$8~pc at the assumed distance, which is
far larger than the Orion Nebula Cluster and somewhat larger than NGC\,6357
\citep{jwang07a}.

The other feature is a deviation from the monotonic trend at $\sim$4\arcmin. A similar
excess can also be found in the identified and unidentified X-ray source profile
(Figs.~\ref{fg:f6}b and c). A $\sim$4\arcmin\ radius partial ring around Wd\,2 is the
most conspicuous feature both in the radio and MIR images (Fig.~\ref{fg:f1}a;
\citealt{whiteoak97,churchwell04}), where gas and dust are compressed by winds and
radiation from massive sources in the central OB association. The possible peak at
$\sim$4\arcmin\ in the X-ray profiles, although the significance is low, suggests that a
second generation of triggered star formation is present in this ring. A deeper X-ray
image would confirm this possibility and identify members of the triggered population.

We identified thirty new OB star candidates based on the NIR photometry of X-ray sources
(\S~\ref{sect:s5-4}). It is interesting to note that many of them are located outside of
the cluster core. In addition to the three off-core early-type stars (WR\,20a, 20b, and
MSP\,18 at $\sim$0\farcm6, $\sim$3\farcm7, $\sim$1\farcm0\ from the Wd\,2 center,
respectively; Fig.~\ref{fg:f1}a) considered to be physically associated with RCW\,49, we
have 14 early-type candidates beyond 3$\arcmin$ of the cluster center. One of them
(HD\,90273) is a spectroscopically-identified O star \citep{cannon19}. This implies
either that these are run-away O stars as suggested for MSP\,18 \citep{uzpen05} or that
the cluster was formed without primordial mass segregation. We note that the OB stars in
the Rosette Nebula Cluster similarly do not exhibit mass segregation
\citep{jwang07b}. Follow-up studies are mandatory to confirm the early-type nature of
these off-core candidates and to understand the cause of their large spatial
distribution.

\subsection{Discrimination of Intrinsic Cluster Members}\label{sect:s5-6}
We detected 10,540 and 9,768 sources in the SIRIUS images on and off of the RCW\,49
region (\S~\ref{sect:s3}; hereafter called the object and the control fields). This
indicates that the NIR sample is seriously contaminated by field stars. Individual
cluster members, most of which are low-mass pre--main-sequence sources, are
traditionally identified by infrared excess emission from their circumstellar disks and
envelopes. MIR excess studies are more sensitive than NIR excess studies
\citep{haisch01}. \citet{whitney04} identified 256 cluster members in the
\textit{Chandra} field from the MIR excess signature based on the MIR color-color
diagrams using \textit{Spitzer}/IRAC. However, the infrared excess is only sensitive to
classical T Tauri stars and protostars with rich circumstellar matter. Weak-line T Tauri
stars are usually missed. On the contrary, X-ray emission in the $\gtrsim$1~keV band
does not significantly depend on whether pre--main-sequence sources have disks or not
\citep{feigelson07}. We now use the X-ray emission as a signature of cluster membership
and examine how the \textit{Chandra} data complement the \textit{Spitzer} data.

For this purpose, we construct three \ksi-band luminosity functions (KLFs). One is the
intrinsic KLF of RCW\,49 obtained by subtracting the KLF of the control field from that
of the object field. The other two are the KLFs of SIRIUS sources with X-ray and MIR
excess emission signatures. We examine what fraction of the first KLF is recovered by
each and the combination of the latter two KLFs.

\begin{figure}[hbtp]
 \figurenum{11}
 \epsscale{1.0}
 \plotone{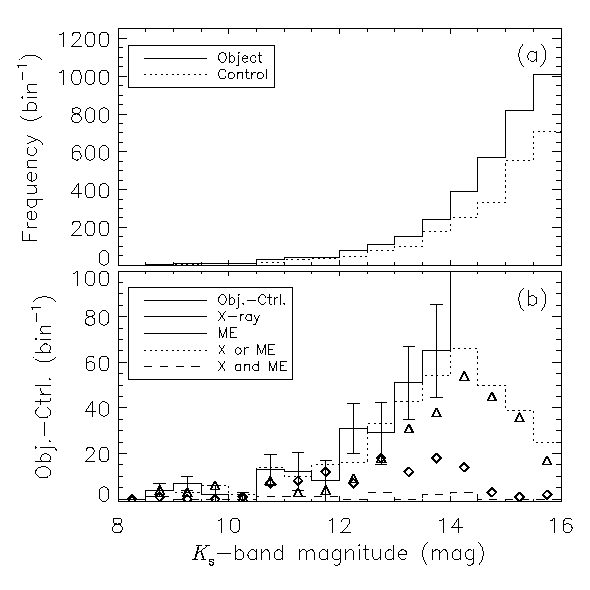}
 \caption{(a) KLFs of the object and control fields from SIRIUS (solid and dashed
 histograms). The KLF of the object field is shifted leftward by 0.1~mag to compensate
 for extra reddening relative to the control field. (b) Intrinsic KLF of RCW\,49 (solid
 histogram) derived as the subtraction of the KLFs of the object and control fields. The
 Poisson uncertainty is given by error bars. The KLFs of SIRIUS sources with X-ray and
 MIR excess (ME) identifications are given by triangles and diamonds, respectively. The
 KLFs of SIRIUS sources with either of and both of the two identifications (the union
 and intersection) are given by dotted and dashed histograms,
 respectively.}\label{fg:f11}
\end{figure}

Figure \ref{fg:f11} (a) shows the KLFs in the object and the control fields
(\textit{solid and dashed histograms}), while the difference of the two is plotted by a
solid histogram in Figure \ref{fg:f11} (b). Sources in the two fields have slightly
different characteristic extinction, for which we compensate by shifting the object KLF
brightward by 0.1~mag based on the following argument. The difference between the mean
NIR colors in the object and control fields represents the difference in the
extinction. In order to avoid the effect of excess emission seen among
pre--main-sequence sources, we use only the (\ji--\hi) color of sources without any
evidence of infrared excess. The mean (\ji--\hi) values in the object and control fields
are 0.94 and 0.83 mag with $\sim$0.01~mag uncertainty. The difference can be converted
to $\Delta A_{K} \sim 0.1$~mag \citep{tokunaga00,benjamin03}.

First, we compare the intrinsic KLF of RCW\,49 with each of the KLFs of sources with
X-ray and MIR excess emission, which are respectively shown with triangles and diamonds
in Figure~\ref{fg:f11} (b). Each of the X-ray and MIR-excess KLFs represent part of the
intrinsic KLF. The recovery rate drops at \ksi~$\gtrsim$~14~mag because the
\textit{Chandra} and \textit{Spitzer} sensitivities do not match that of SIRIUS. Also,
the \textit{Spitzer} image suffers confusion in the most crowded part of the
cluster. The recovery of cluster members by the X-ray KLF is also expected to drop in
the magnitude range for late B and A type sources, because some of them are
intrinsically X-ray faint unlike earlier and later type sources
\citep{preibisch05}. These sources would appear at $\sim$12~mag at the assumed
distance. The smaller X-ray recovery rate at 12.0--12.5~mag may be attributable to this
effect (Fig.~\ref{fg:f11}b).

We nonetheless do not attempt to correct statistically for these types of incompleteness
because we are interested in how many of the intrinsic cluster members are
\textit{individually} identified by the current data set. The total counts of sources in
the intrinsic KLF at \ki$<$14.0 and $<$14.5~mag are 221 and 358. Respectively, 125
(57\%) and 179 (50\%) of them are recovered in the X-ray KLF, while 84 (38\%) and 98
(27\%) are in the MIR excess KLF. Each of the X-ray and MIR excess KLFs represent about
30--60\% of the intrinsic RCW\,49 KLF.

Second, we compare the intrinsic KLF with the combination of X-ray and MIR excess
KLFs. The KLFs of the union and the intersection sets of X-ray and MIR excess sources
are shown by dotted and dashed histograms in Figure~\ref{fg:f11} (b). The union set
recovers 90\% (200) and 74\% (266) of the intrinsic cluster members at \ki$<$14.0 and
$<$14.5~mag, respectively. The intersection set is very small. These indicate that X-ray
and MIR excess emission selections operate in a complementary fashion in identifying
cluster members and that the combination of the two recovers a large fraction of the
members. The complementarity probably comes from the fact that richer circumstellar
matter around pre--main-sequence sources increases the MIR excess detection rate due to
stronger emission and decreases the X-ray detection rate due to larger extinction. This
is best examined in the NIR color-color diagram in Figure~\ref{fg:f10} (b). SIRIUS
sources with X-ray and MIR excess signatures (blue and red dots) are segregated in the
plot. The X-ray emission preferentially detects sources with no or smaller NIR excess.

\subsection{X-ray Emission from Massive Stars}\label{sect:s5-7}
We present the results of detailed X-ray spectroscopy for five exceptionally bright
X-ray sources with more than 200 counts. Two of them are Wolf-Rayet stars (WR\,20a and
20b) and the rest are O-type stars (MSP\,18, 167, and 188). All these sources are
located at the top right end of X-ray and NIR brightness plots (Fig.~\ref{fg:f9}).

\subsubsection{Wolf-Rayet stars}\label{sect:s5-7-1}
\begin{figure*}[hbtp]
 \figurenum{12}
 \epsscale{1.0}
 \plotone{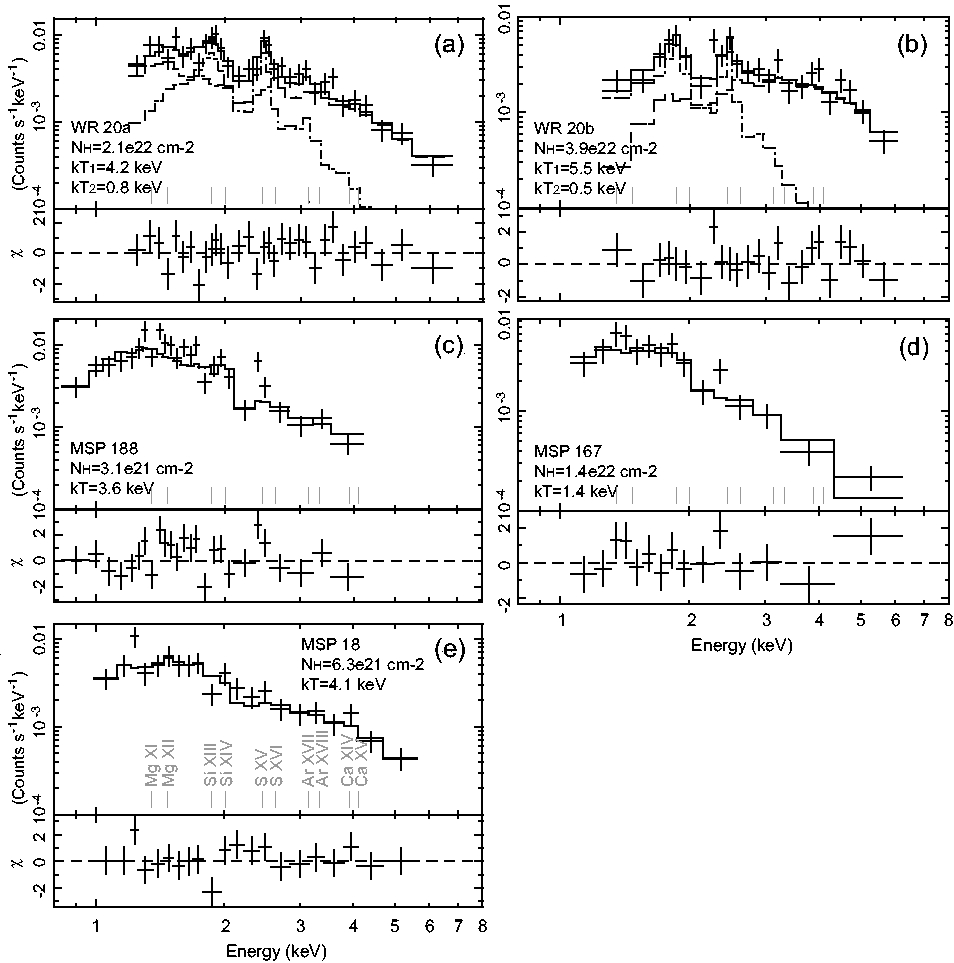}
 \caption{X-ray spectra of bright sources. The source identifications and the models
 employed to fit the spectra are given in Table~\ref{tb:t4}. Grouped data with errors
 are shown in the upper panels over-plotted with the best-fit model convolved with
 mirror and detector responses (solid line). The energy of the K$\alpha$ lines of some
 representative highly ionized ions are shown. The lower panels show the residuals of
 the fit.}\label{fg:f12}
\end{figure*}

The X-ray spectra of WR\,20a and WR\,20b are shown in Figure~\ref{fg:f12} (a) and (b),
respectively. Both of them are characterized by emission lines from highly ionized ions
such as Mg, Si, S, Ar, and Ca, and a hard tail up to $\sim$5~keV. The \ion{Si}{13}
K$\alpha$ emission as well as the hard tail indicate that the emission is more likely
from a plasma of multiple temperatures than from that of a uniform temperature. We
present the results of a two-temperature plasma fit in Table~\ref{tb:t4}. In the
fitting, the abundances of conspicuous lines (Si and S) were thawed to obtain the
best-fit values. The abundances of other elements were fixed to be 1 solar.  It is known
that hydrogen is heavily depleted in Wolf-Rayet stars (e.g.,
\citealt{vanderhucht86,morris00}), but we used the cosmic abundance \citep{anders89} as
we do not know the level of hydrogen depletion for these stars. Once it is known, the
abundance values presented here should be decreased accordingly.

\begin{figure}[hbtp]
 \figurenum{13}
 \epsscale{1.0}
 \plotone{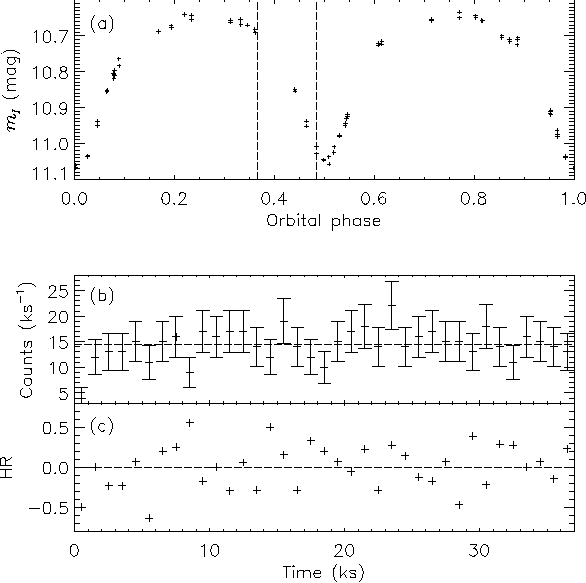}
 \caption{Light curves of the binary system WR\,20a. (a) \textit{I}-band magnitude of
 one orbit \citep{bonanos04}. (b) X-ray count rate in the 0.5--8.0~keV band and (c) the
 hardness ratio (HR) in the phase 0.36 to 0.48 (the range is indicated by two dotted
 lines in a; the uncertainty is about $\pm$0.2). We used the ephemeris by
 \citet{bonanos04} with an orbital period of 3.686 days and a Julian date origin of
 2,453,124.569 day. HR here is defined as (H$-$S)/(H$+$S) where H and S are the count
 rates in the hard (2.0--8.0~keV) and the soft (0.5--2.0~keV) bands, respectively.}\label{fg:f13}
\end{figure}

WR\,20a is an eclipsing binary of two Wolf-Rayet stars of spectral type WN6ha
\citep{bonanos04,rauw05}. Using the best-fit orbital solution by \citet{bonanos04}, the
\textit{Chandra} observation covers the orbital phase from 0.36 to 0.48. Because of the
uncertainty in the orbital period determination of 3.686 $\pm$ 0.01 days, the
\textit{Chandra} phase range has uncertainty from 0.18--0.30 to 0.55--0.67. We compare
the X-ray light curves and the eclipsing optical light curve (Fig.~\ref{fg:f13}) with the
aim to obtain a hint in the X-ray production mechanism of this source. One artificial
effect is introduced by the fact that the source was observed in a gap between ACIS I2
and I3. The source oscillated across the gap at a period of 1000~s due to the dithering
motion of the telescope. About 60\% of the photons were lost in the gap. We binned the
light curves with a binning size of 1000~s to cancel this effect.

No apparent variation was found in either the average X-ray count rate
(Fig.~\ref{fg:f13}b) or the hardness ratio (Fig.~\ref{fg:f13}c). In contrast, a clear
decline is seen in the \textit{I}-band magnitude (Fig.~\ref{fg:f13}a) during the phase
of about 0.3--0.7. The \textit{Chandra} phase coverage is uncertain from 0.18--0.30 to
0.55--0.67. When its lower limit is adopted, the \textit{Chandra} observation was
conducted out of eclipses. When other values are adopted, it was conducted during an
optical eclipse. In this case, the lack of the eclipse in the X-ray light curves
indicates that the X-ray emitting region is far enough from the stellar surface so as
not affected by the photospheric eclipsing. From the orbital solution by
\citet{bonanos04}, the radii of the two stars are $\sim$19$R_{\odot}$ and $a\sin{(i)}$
is $\sim$53$R_{\odot}$, where $a$ is the semi-major axis of the orbit,
$i=$74.5$^{\circ}$ is the inclination angle, and $R_{\odot}$ is the radius of the
Sun. The system is observed close to the edge-on in the line of sight. At the middle of
the eclipse, the two stars are separated by $2a\cos{(i)} \sim 29 R_{\odot}$. Therefore,
the X-ray emitting region is extended to at least a few stellar radii.

\subsubsection{O-type Stars}\label{sect:s5-7-2}
At least two out of three bright O stars show emission line features of highly ionized
Mg, Si and S, indicating their thermal plasma origin. We fit all three spectra with
a thin-thermal plasma model with variable abundance of elements with
conspicuous features. Table~\ref{tb:t4} gives the results. All of them are characterized
by a hard spectrum with the best-fit plasma temperature of $\gtrsim$1.5~keV and luminous
X-ray brightness of 2--8$\times$10$^{32}$~ergs~s$^{-1}$.

Such hard and luminous X-ray-emitting O stars have been reported in some other massive
SFRs \citep{albacete03,rho04,skinner05,broos07,wolk06}. The X-ray properties of these
sources are quite different from the classical view of main-sequence O-type stars
characterized by soft thermal spectra of $\lesssim$1~keV and the X-ray to bolometric
luminosity ratio of $\sim$10$^{-7}$ \citep{berghoefer97}. Hard X-ray--emitting O stars
in massive SFRs should be a new distinct class of X-ray sources with a different X-ray
production mechanism.

Several interpretations have been proposed for such hard X-ray emission from O
stars. One is magnetically-confined wind shocks
\citep{babel97,schulz03,gagne05,stelzer05}, where magnetized winds produce hard X-ray
emission in large-scale shocks near the stellar surface. Another is similar to the X-ray
emission in WR binaries; i. e., winds from each component of a close binary of two
O-type stars collide with each other to make hard X-rays
\citep{zhekov00,skinner07}. The hard X-rays might be from the magnetic reconnections
between the two binary components \citep{schulz06}. The binary nature of the three
O-type stars in RCW\,49 is not confirmed yet, but these sources might be
``X-ray-spectroscopic binaries'' if this interpretation is correct. Yet another idea is
that inverse Compton emission of UV photons from these O stars accounts for their hard
spectra \citep{albacete03}. The X-ray production mechanism of such curious O stars is
not yet certain, but enriching the sample using plots like Figure~\ref{fg:f9} may lead
to a better understanding of the mechanism.

\section{Summary}\label{sect:s6}
We observed the Galactic massive star-forming region RCW\,49 and its central OB
association Westerlund 2 using \textit{Chandra}/ACIS, IRSF/SIRIUS, and
\textit{Spitzer}/IRAC. The main results are summarized as follows.

\begin{enumerate}
 \item 468 X-ray sources were detected in a $\sim$40~ks ACIS exposure covering a
 $\sim$17\arcmin$\times$17\arcmin\ field. Previous X-ray studies detected less than 10
 sources in the studied field and almost all the X-ray sources reported here are new
 detections.  The results of X-ray photometric, timing, and spectroscopic analyses are
 presented and summarized in Tables~\ref{tb:t1} and \ref{tb:t2}.

 \item We obtained SIRIUS images in the \ji, \hi, and \ksi\ bands of a concentric
 $\sim$8\farcm3$\times$8\farcm3 field and extracted 10,540 sources at a 10~$\sigma$
 detection limits of $\sim$19.0 (\ji), $\sim$17.8 (\hi), and $\sim$16.8 (\ksi) mag.

 \item 379 X-ray sources were identified with optical, NIR, and MIR counterparts using
 NOMAD, 2MASS, SIRIUS, and GLIMPSE data. The results are given in Table~\ref{tb:t3}.

 \item The central OB association Westerlund 2 was resolved in the X-ray image for the
 first time. X-ray emission was detected from all of the spectroscopically-identified
 early-type stars consisting of two Wolf Rayet stars and a dozen O type stars.

 \item About 86\% of the X-ray sources with optical or infrared identifications were
 found to be cluster members by comparing with the X-ray population in a control field
 in the Galactic Plane. The X-ray overpopulation against the control field is seen in
 all measured flux ranges.

 \item We compared the intrinsic \ki-band luminosity function of NIR sources in RCW\,49
 with the KLFs of sources with X-ray or MIR excess identification. We found that
 30--60\% of the cluster members at \ksi$<$14~mag are individually identified using the
 two observational signatures of cluster membership. When X-ray and MIR excess
 signatures are combined, about 90\% of cluster sources are individually located. The
 two indicators of the cluster membership work in a complementary fashion, in which the
 former excels in detecting weak-line T Tauri stars with smaller attenuation while the
 latter predominantly detects classical T Tauri stars or protostars with richer
 circumstellar matter.

 \item A loose constraint on the distance to RCW\,49 was derived to be 2--5~kpc based on
 the mean hard-band X-ray luminosity of T Tauri stars in the mass range of
 2.0--2.7~$M_{\odot}$.

 \item The cluster X-ray population consists of low-mass pre--main-sequence sources and
 early-type sources based on X-ray and NIR photometry. Late-type pre--main-sequence
 sources, Wolf-Rayet stars, and early-type main sequence stars occupy different regions
 of X-ray versus NIR brightness plots. Thirty new OB candidates were identified based on
 the NIR color-magnitude diagram, several times more than the number of
 optically-identified OB stars.

 \item Using the X-ray surface number density radial profiles, we gave an estimate of
 the cluster radius to be 6\arcmin--7\arcmin, or $\sim$8~pc at the assumed distance. A
 possible excess is seen at $\sim$4\arcmin\ of the center, suggesting that secondary
 star formation is taking place at the ring. Fourteen OB star candidates are found
 outside of Wd\,2, suggesting an absence of mass segregation.

 \item Detailed X-ray spectroscopy of the two Wolf-Rayet stars was presented. Both
 sources show two-temperature thermal plasmas with characteristic temperatures of
 0.5--1.0 and 4--6~keV and 1--8~keV luminosities of 2$\times$10$^{33}$~ergs~s$^{-1}$. No
 X-ray flux or hardness variation was found from WR\,20a, an eclipsing binary,
 suggesting that the X-ray emission arises far from stellar surfaces.

 \item Detailed X-ray spectroscopy of three luminous O-type stars were presented. They
 are characterized by thermal spectra with a temperature of $\gtrsim$1.5~keV. These are
 among the growing samples of hard X-ray emitting O stars recently found in some massive
 star-forming regions. Early-type stars appears to have two sub groups, hard and soft,
 based on the X-ray spectroscopy.
\end{enumerate}

\acknowledgments

The authors express gratitude to Barbara Whitney for providing the \textit{Spitzer}
data. M.\,T. thanks for the hospitality at the Academia Sinica Institute of Astronomy
and Astrophysics in Taipei and Theoretical Institute of Advanced Research in
Astrophysics in Hsinchu, Taiwan during the course of this work. We wish to thank the
staff of the South African Astronomical Observatory for their kind support during our
observation.

This work is financially supported by the Japan Society for the Promotion of Science,
the NASA contract NAS8-38252, and the Chandra Contract SV4-74018 issued by the
\textit{Chandra} X-ray Center on behalf of NASA under Contract NAS8-08060. The
IRSF/SIRIUS project was financially supported by the Sumitomo Foundation and the
Grant-in-Aids for Scientific Research Numbers 10147207 and 10147214 of Ministry of
Education, Culture, Sports, Science, and Technology of Japan.

This work is based in part on observations made with the \textit{Spitzer Space
Telescope} operated by Jet Propulsion Laboratory under a contract with NASA and data
products from the Two Micron All Sky Survey, which is a joint project of the University
of Massachusetts and the Infrared Processing and Analysis Center/California Institute of
Technology, funded by the National Aeronautics and Space Administration and the National
Science Foundation.

We also made use of the SIMBAD database, operated at CDS, Strasbourg, France. IRAF is
distributed by the National Optical Astronomy Observatories, which are operated by the
Association of Universities for Research in Astronomy, Inc., under cooperative agreement
with the National Science Foundation.

Facilities: \facility{CXO(ACIS)}, \facility{IRSF(SIRIUS)}, \facility{SST(IRAC)}

\clearpage

\begin{turnpage}
\begin{deluxetable}{rcrrrrrrrrrrrrrcccccc}
\centering
\tabletypesize{\scriptsize}
\tablewidth{0pt}
\tablecolumns{21}
\tablecaption{ {\it Chandra} Catalog: Basic Source Properties \label{tb:t1}}
\tablehead{
 \multicolumn{2}{c}{Source} &
 &
 \multicolumn{4}{c}{Position} &
 &
 \multicolumn{5}{c}{Extracted Counts} &
 &
 \multicolumn{7}{c}{Characteristics} \\
 \cline{1-2} \cline{4-7} \cline{9-13} \cline{15-21}
 \colhead{Seq} & \colhead{CXOU J} &
 & 
 \colhead{$\alpha_{\rm J2000}$} & \colhead{$\delta_{\rm J2000}$} & \colhead{Err} & \colhead{$\theta$} &
 & 
 \colhead{$C_{\rm{net}}$} & \colhead{$\Delta C_{\rm{net}}$} & \colhead{$C^{\prime}_{\rm{bkg}}$} & \colhead{$C_{\rm{net,hard}}$} & \colhead{PSF} &   
 &
 \colhead{PS} & \colhead{$\log P_B$} & \colhead{Anom} & \colhead{Var} & \colhead{EffExp} & \colhead{$E_{\rm{median}}$}  & \colhead{Photo \fx}\\
 \colhead{\#} & \colhead{} &
 & 
 \colhead{(deg)} & \colhead{(deg)} & \colhead{(\arcsec)} & \colhead{(\arcmin)} &
 & 
 \colhead{} & \colhead{} & \colhead{} & \colhead{} & 
 \colhead{Frac} &
 &
 \colhead{} & \colhead{} & \colhead{} & \colhead{} & \colhead{(ks)} & \colhead{(keV)} & \colhead{(ergs~s~cm$^{-2}$)}
 \\
 \colhead{(1)} & \colhead{(2)} &
 & 
 \colhead{(3)} & \colhead{(4)} & \colhead{(5)} & \colhead{(6)} &
 & 
 \colhead{(7)} & \colhead{(8)} & \colhead{(9)} & \colhead{(10)} & \colhead{(11)} &
 & 
 \colhead{(12)} & \colhead{(13)} & \colhead{(14)} & \colhead{(15)} & \colhead{(16)} & \colhead{(17)} & \colhead{(18)}
}

\startdata
1 & 102257.09$-$574403.0 & & 155.73789 & $-$57.73418 & 1.2 & 8.6 & & 7.8 & 4.2 & 5.2 & 0.0 & 0.89 & & 1.6 & $-$2.5 & .... & a & 20.7 & 1.0 & 2.8e$-$16 \\
2 & 102259.75$-$574157.1 & & 155.74899 & $-$57.69922 & 0.7 & 8.8 & & 28.8 & 6.3 & 4.2 & 0.0 & 0.90 & & 4.2 & $<-$5 & .... & a & 20.1 & 0.9 & 4.3e$-$15 \\
3 & 102300.48$-$574226.1 & & 155.75202 & $-$57.70726 & 0.5 & 8.5 & & 53.3 & 8.1 & 3.7 & 2.7 & 0.90 & & 6.2 & $<-$5 & .... & a & 20.9 & 1.0 & 9.9e$-$15 \\
4 & 102300.49$-$574307.5 & & 155.75208 & $-$57.71875 & 1.1 & 8.3 & & 7.9 & 3.9 & 3.1 & 0.0 & 0.90 & & 1.8 & $-$3.4 & .... & a & 19.9 & 1.2 & 1.6e$-$15 \\
5 & 102302.84$-$574606.9 & & 155.76185 & $-$57.76859 & 0.3 & 7.7 & & 110.4 & 11.4 & 8.6 & 73.7 & 0.90 & & 9.2 & $<-$5 & .... & a & 21.2 & 2.8 & 5.4e$-$14 \\
6 & 102305.27$-$574150.6 & & 155.77196 & $-$57.69741 & 0.9 & 8.1 & & 9.1 & 4.0 & 2.9 & 8.0 & 0.89 & & 2.0 & $-$4.2 & .... & a & 20.5 & 3.0 & 5.4e$-$15 \\
7 & 102306.68$-$574353.2 & & 155.77787 & $-$57.73145 & 0.3 & 7.3 & & 77.1 & 9.4 & 1.9 & 8.8 & 0.89 & & 7.8 & $<-$5 & .... & c & 21.5 & 1.1 & 1.6e$-$14 \\
8 & 102308.67$-$574631.2 & & 155.78616 & $-$57.77535 & 0.7 & 7.0 & & 10.5 & 4.1 & 2.5 & 0.4 & 0.90 & & 2.2 & $<-$5 & g... & \nodata & 19.7 & 1.2 & 2.4e$-$15 \\
9 & 102309.23$-$574620.4 & & 155.78848 & $-$57.77235 & 0.9 & 6.9 & & 6.8 & 3.5 & 2.2 & 0.0 & 0.90 & & 1.6 & $-$3.3 & g... & \nodata & 19.8 & 0.9 & 3.8e$-$16 \\
10 & 102309.82$-$574942.2 & & 155.79096 & $-$57.82841 & 0.5 & 8.1 & & 35.5 & 6.7 & 2.5 & 20.3 & 0.90 & & 4.9 & $<-$5 & .... & b & 18.6 & 2.2 & 1.5e$-$14 \\
\enddata
\tablecomments{
{\bf Column 1:} X-ray catalog sequence number, sorted by R.\,A.
{\bf Column 2:} IAU designation.
{\bf Columns 3, 4:} R.\,A. and decl. for the equinox J2000.0.
{\bf Column 5:} Estimated standard deviation of the random component of the position
error, $\sqrt{\sigma_x^2 + \sigma_y^2}$.  The single-axis position errors, $\sigma_x$
and $\sigma_y$, are estimated from the single-axis standard deviations of the PSF inside
the extraction region and the number of counts extracted.
{\bf Column 6:} Off-axis angle.
{\bf Columns 7, 8:} Estimated net counts extracted in the total energy band (0.5--8~keV);
average of the upper and lower 1~$\sigma$ errors on column 7.
{\bf Column 9:} Background counts scaled to the source extraction area (total band).
{\bf Column 10:} Estimated net counts extracted in the hard energy band (2--8~keV).
{\bf Column 11:} Fraction of the PSF (at 1.497 keV) enclosed within the extraction
region. A reduced PSF fraction (significantly below 90\%) may indicate that
the source is in a crowded region.
{\bf Column 12:} Photometric significance.
{\bf Column 13:} Logarithmic probability that extracted counts (total band) are solely
from background. 
{\bf Column 14:} Source anomalies: g = fractional time that source was on a detector
(FRACEXPO from {\em mkarf}) is $<0.9$ ; e = source on field edge; p = source piled up; s
= source on readout streak.
{\bf Column 15:} Variability characterization based on K-S statistic (total band): a =
no evidence for variability ($0.05<P_{KS}$); b = possibly variable
($0.005<P_{KS}<0.05$); c = definitely variable ($P_{KS}<0.005$).  No value is reported
for sources in chip gaps or on field edges.
{\bf Column 16:} Effective exposure time: approximate time the source would have to be
observed on-axis to obtain the reported number of
counts.
{\bf Column 17:} Background-corrected median photon energy (total band).
{\bf Column 18:} Photometric flux estimate in the 0.5--8.0~keV band. See discussion in \S~\ref{sect:s4-1-3}.
}
\tablecomments{Table \ref{tb:t1} is published in its entirety in the electronic edition of the {\it Astrophysical Journal}.  A portion is shown here for guidance regarding its form and content.}
\end{deluxetable}

\end{turnpage}

\clearpage

\begin{turnpage}
\begin{deluxetable}{rcrrrcccrccccc}
 \centering
 \tabletypesize{\small}
 \tablewidth{0pt}
 \tablecolumns{14}
 \tablecaption{X-ray Spectroscopic Fits\label{tb:t2}}
 \tablehead{
 \multicolumn{4}{c}{Source\tablenotemark{a}} &
 &
 \multicolumn{3}{c}{Spectral Fit\tablenotemark{b}} &
 &
 \multicolumn{5}{c}{X-ray Luminosities\tablenotemark{c}} \\
 \cline{1-4} \cline{6-8} \cline{10-14}
 \colhead{Seq} & \colhead{CXOU J} & \colhead{$C_{\rm{net}}$} & \colhead{PS} &
 &
 \colhead{$\log \nh$} & \colhead{\kt} & \colhead{$\log \emm$} &  
 &
 \colhead{$\log L_{\rm{s}}$} & \colhead{$\log L_{\rm{h}}$} & \colhead{$\log L_{\rm{h,c}}$} & \colhead{$\log L_{\rm{t}}$} & \colhead{$\log L_{\rm{t,c}}$} \\
 \colhead{\#} & \colhead{} & \colhead{} & \colhead{} &
 &
 \colhead{(cm$^{-2}$)} & \colhead{(keV)} & \colhead{(cm$^{-3}$)} & 
 &
 \multicolumn{5}{c}{(ergs s$^{-1}$)} \\
 \colhead{(1)} & \colhead{(2)} & \colhead{(3)} & \colhead{(4)} &
 &
 \colhead{(5)} & \colhead{(6)} & \colhead{(7)} &
 &
 \colhead{(8)} & \colhead{(9)} & \colhead{(10)} &\colhead{(11)} & \colhead{(12)}
}
\startdata
5 & 102302.84$-$574606.9 & 110.4 & 9.2 & & {\tiny $-0.2$} 22.3 {\tiny $+0.2$} & {\tiny $-2.1$} 4.3 {\tiny $+5.4$} & 55.3 & & 31.07 & 32.04 & 32.13 & 32.09 & 32.39 \\
29 & 102328.78$-$574629.3 & 106.5 & 9.4 & & 22.5 & {\tiny $-0.5$} 1.6 {\tiny $+0.7$} & {\tiny $-0.3$} 55.7 {\tiny $+0.4$} & & 31.10 & 31.92 & 32.13 & 31.98 & 32.68 \\
68 & 102347.43$-$574755.8 & 157.7 & 11.6 & & {\tiny $-0.1$} 22.2 {\tiny $+0.1$} & {\tiny $-3.9$} 7.4 \phantom{{\tiny $-3.9$}} & {\tiny $-0.1$} 55.3 \phantom{{\tiny $-0.1$}} & & 31.18 & 32.19 & 32.26 & 32.23 & 32.46 \\
132 & 102358.01$-$574548.8 & 532.6 & 22.1 & & {\tiny $-0.08$} 22.3 {\tiny $+0.06$} & {\tiny $-0.3$} 1.9 {\tiny $+0.5$} & {\tiny $-0.1$} 56.4 {\tiny $+0.1$} & & 32.12 & 32.80 & 32.94 & 32.89 & 33.39 \\
192 & 102401.19$-$574531.0 & 451.9 & 20.3 & & {\tiny $-0.1$} 21.8 {\tiny $+0.1$} & {\tiny $-0.4$} 2.0 {\tiny $+0.4$} & {\tiny $-0.08$} 55.7 {\tiny $+0.11$} & & 31.96 & 32.14 & 32.18 & 32.36 & 32.62 \\
217 & 102402.06$-$574527.9 & 246.2 & 14.7 & & {\tiny $-0.1$} 22.0 {\tiny $+0.1$} & {\tiny $-0.5$} 2.0 {\tiny $+0.7$} & {\tiny $-0.1$} 55.5 {\tiny $+0.2$} & & 31.64 & 31.99 & 32.05 & 32.15 & 32.48 \\
229 & 102402.42$-$574436.1 & 352.7 & 17.8 & & {\tiny $-0.1$} 21.8 {\tiny $+0.1$} & {\tiny $-1.1$} 4.2 {\tiny $+2.5$} & {\tiny $-0.10$} 55.8 {\tiny $+0.10$} & & 32.12 & 32.63 & 32.66 & 32.75 & 32.92 \\
396 & 102418.39$-$574829.8 & 398.4 & 19.0 & & {\tiny $-0.09$} 22.4 {\tiny $+0.08$} & {\tiny $-0.7$} 3.0 {\tiny $+1.3$} & {\tiny $-0.1$} 56.0 {\tiny $+0.2$} & & 31.56 & 32.56 & 32.69 & 32.60 & 33.02 \\
463 & 102454.55$-$574842.9 & 193.9 & 12.9 & & 22.5 & \nodata & 55.7 & & \nodata & \nodata & \nodata & \nodata & \nodata \\
\enddata

\tablenotetext{a}{ For convenience, {\bf columns 1--4} reproduce the source
identification, net counts, and photometric significance data from Table~\ref{tb:t1}.}
\tablenotetext{b}{Sources with photometric significance of larger than 2 were fit with
an absorbed thin-thermal plasma model. The abundance is fixed to be 0.3 times solar
value. {\bf Columns 5 and 6} present the best-fit values for the extinction column density and
plasma temperature parameters.  {\bf Column 7} presents the emission measure for the
model spectrum, assuming a distance of 4.2~kpc. Uncertainties represent 90\% confidence
intervals. More significant digits are used for uncertainties $<=0.1$ in order to avoid
large rounding errors; for consistency, the same number of significant digits is used
for both lower and upper uncertainties. Uncertainties are missing when XSPEC was unable
to compute them or when their values were so large that the parameter is effectively
unconstrained.  Fits lacking uncertainties, fits with large uncertainties, and fits with
frozen parameters should be viewed merely as splines to the data to obtain rough
estimates of luminosities; the listed parameter values are unreliable.}
\tablenotetext{c}{X-ray luminosities are presented in {\bf columns 8--12}: s = soft band
(0.5--2 keV); h = hard band (2--8 keV); t = total band (0.5--8 keV).
Absorption-corrected luminosities are subscripted with a ``c''; they are omitted when
$\log N_H > 22.5$ since the soft band emission is essentially unmeasurable. Luminosities
are derived assuming a distance of 4.2~kpc.}
\tablecomments{Table \ref{tb:t2} is published in its entirety in the electronic edition of the {\it Astrophysical Journal}.  A portion is shown here for guidance regarding its form and content.}
\end{deluxetable}

\end{turnpage}

\clearpage

\begin{turnpage}
\begin{deluxetable}{rccccrrrrrrrrl}
 \centering
 \tabletypesize{\footnotesize}
 \tablecaption{Optical, NIR, and MIR Counterparts of ACIS Sources\label{tb:t3}}
 \tablecolumns{14}
 \tablewidth{0pt}
 \tablehead{
 \colhead{X-ray\tablenotemark{a}} &
 \colhead{NOMAD} &
 \colhead{2MASS} &
 \colhead{IRAC} &
 \colhead{NIR\tablenotemark{b}} &
 \colhead{\ri} &
 \colhead{\ji} &
 \colhead{\hi} &
 \colhead{\ksi} &
 \colhead{[3.6]} &
 \colhead{[4.5]} &
 \colhead{[5.8]} &
 \colhead{[8.0]} &
 \colhead{Flags\tablenotemark{c}} \\
 \colhead{ID} &
 \colhead{ID} &
 \colhead{ID} &
 \colhead{ID} &
 \colhead{flag} &
 \colhead{(mag)} &
 \colhead{(mag)} &
 \colhead{(mag)} &
 \colhead{(mag)} &
 \colhead{(mag)} &
 \colhead{(mag)} &
 \colhead{(mag)} &
 \colhead{(mag)} &
 \colhead{\& IDs}
 }
 \startdata
  3 & \nodata & \nodata & G284.1287$-$00.3588 & \nodata & \nodata & \nodata & \nodata & \nodata & 11.0 & 10.9 & 10.9 & 10.9 & BM22\\
 63 & 0323$-$0248919 & 10234445$-$5738316 & G284.1762$-$00.2515 & T & 9.0 & 8.6 & 8.6 & 8.6 & 8.5 & 8.6 & 8.6 & 8.5 & HD90273, BM5, ET\\
112 & 0322$-$0246800 & 10235617$-$5745299 & G284.2604$-$00.3358 & S & 13.8 & 10.6 & 10.1 & 9.8 & 9.4 & 9.4 & 9.4 & 9.7 & MSP182, ET\\
132 & 0322$-$0246854 & 10235800$-$5745489 & G284.2667$-$00.3380 & T & 12.6 & 8.9 & 8.1 & 7.6 & 7.1 & 6.8 & 6.7 & 6.3 & WR20a, BM25, ET, ME\\
171 & \nodata & \nodata & G284.2676$-$00.3293 & S & \nodata & 10.7 & 10.1 & 9.8 & 9.4 & 9.4 & 9.2 & \nodata & MSP151, ET\\
172 & 0322$-$0246947 & 10240049$-$5744444 & G284.2618$-$00.3199 & S & 11.1 & 12.0 & 11.4 & 11.1 & 10.7 & 10.7 & 10.8 & \nodata & MSP44, ET\\
181 & 0322$-$0246954 & 10240073$-$5745253 & G284.2684$-$00.3293 & T & \nodata & \nodata & \nodata & \nodata & 9.2 & 9.2 & 9.3 & 9.1 & MSP157\\
192 & 0322$-$0246983 & 10240125$-$5745308 & G284.2701$-$00.3299 & S & \nodata & 9.7 & 9.1 & 8.9 & 8.3 & 8.2 & 8.1 & \nodata & MSP188, ET\\
202 & 0322$-$0246993 & 10240151$-$5745569 & G284.2744$-$00.3358 & S & 14.3 & 10.6 & 9.9 & 9.6 & 9.2 & 9.2 & 9.2 & \nodata & MSP263, ET\\
217 & 0322$-$0247016 & 10240201$-$5745279 & G284.2711$-$00.3283 & S & \nodata & 10.3 & 9.7 & 9.4 & 8.7 & 8.6 & 8.5 & \nodata & MSP167, ET\\
224 & 0322$-$0247034 & 10240230$-$5745351 & G284.2726$-$00.3297 & S & \nodata & 10.5 & 9.2 & 9.1 & 8.3 & 8.3 & 8.2 & \nodata & MSP203, ET\\
226 & 0322$-$0247036 & 10240237$-$5745307 & G284.2722$-$00.3286 & S & \nodata & 9.8 & 9.2 & 8.9 & 8.2 & 8.1 & 7.9 & \nodata & MSP183, ET\\
229 & 0322$-$0247041 & 10240243$-$5744359 & G284.2642$-$00.3156 & S & 12.2 & 9.6 & 9.1 & 9.0 & 8.2 & 7.9 & 7.2 & 4.8 & MSP18, BM23, ET, ME\\
240 & \nodata & \nodata & G284.2732$-$00.3291 & S & \nodata & 10.7 & 10.1 & 9.9 & 9.3 & 9.3 & 9.2 & \nodata & MSP199, ET\\
298 & 0322$-$0247156 & 10240489$-$5745282 & G284.2766$-$00.3250 & T & 11.9 & 10.5 & 10.0 & 9.6 & 9.4 & 9.3 & 9.4 & \nodata & MSP171, ET\\
389 & 0323$-$0249498 & 10241665$-$5739310 & G284.2458$-$00.2271 & T & 15.9 & 13.5 & 12.9 & 12.6 & 12.3 & 12.2 & 11.7 & \nodata & BM16\\
396 & 0321$-$0235221 & 10241839$-$5748297 & G284.3288$-$00.3517 & T & 12.0 & 8.7 & 7.8 & 7.2 & 6.8 & 6.1 & 5.8 & 5.5 & WR20b, ET\\
466 & 0322$-$0248312 & 10245850$-$5747379 & G284.3965$-$00.2921 & T & 14.0 & 13.2 & 12.5 & 12.4 & 12.1 & 12.1 & 11.7 & \nodata & BM9\\
  \enddata
  \tablenotetext{a}{Source sequence numbers follow Table~\ref{tb:t1}.}
  \tablenotetext{b}{``T'' and ``S'' indicate that the NIR photometry is from 2MASS and
  SIRIUS data, respectively. NIR magnitudes are given only for good photometry; i. e.,
  flags A, B, or C for 2MASS sources and the uncertainty of $<$0.1~mag for SIRIUS sources.}
  \tablenotetext{c}{Three types of flags (``NE'', ``ME'', and ``ET'') indicate the NIR
  excess sources identified in the NIR color-color diagram (Fig.~\ref{fg:f9}a), MIR
  excess sources \citep{whitney04}, and early-type star candidates that have a brighter
  \ki-band magnitude than a main-sequence B2V star in the NIR color-magnitude diagram
  (Fig.~\ref{fg:f9}b). Also given are the names in the literature (MSP from
  \citet{moffat91} and BM from \citet{belloni94}).}
  \tablecomments{The complete version of this table is in the electronic edition of the
  Journal. The printed edition contains only a sample.}
\end{deluxetable}

\end{turnpage}

\clearpage

\begin{turnpage}
\begin{deluxetable}{ccccccccccccc}
 \centering
 \tabletypesize{\tiny}
 \tablewidth{0pt}
 \tablecolumns{13}
 \tablecaption{Detailed X-ray Spectroscopy of Bright Sources\label{tb:t4}}
 \tablehead{
 \colhead{X-ray\tablenotemark{a}} & 
 \colhead{Name} & 
 \colhead{Model\tablenotemark{b}} & 
 \colhead{Label\tablenotemark{c}} &
 \colhead{\nh\tablenotemark{d}} & 
 \colhead{\kt$_{1}$\tablenotemark{b,d}} &
 \colhead{\kt$_{2}$\tablenotemark{b,d}} &
 \colhead{\emm$_{1}$/\emm$_{2}$} &
 \colhead{\fx\tablenotemark{e}} &
 \colhead{\lx\tablenotemark{e}} &
 \colhead{Si\tablenotemark{d}} & 
 \colhead{S\tablenotemark{d}} & 
 \colhead{$\chi^{2}$/d.o.f.}\\
 \colhead{ID} &
 &
 &
 &
 \colhead{(10$^{22}$~cm$^{-2}$)} &
 \colhead{(keV)} &
 \colhead{(keV)} &
    &
 \colhead{(ergs s$^{-1}$~cm$^{-2}$)} &
 \colhead{(ergs s$^{-1}$)} &
 \colhead{($Z_{\odot}$)} &
 \colhead{($Z_{\odot}$)} &
 }
 \startdata
 132 & WR\,20a & 1T & \nodata & 2.0 (1.1--2.4) & 2.0 (1.7--3.3) & \nodata & \nodata & 3.7$\times$10$^{-13}$ & 1.5$\times$10$^{33}$ & 0.7 (--2.8) & 1.7 (1.0--5.1) & 40.8/27 \\
 &         & 2T & (a) & 2.1 (0.9--3.0) & 4.2 (2.2--) & 0.8 (0.08--1.1) & 0.6 & 3.9$\times$10$^{-13}$ & 2.2$\times$10$^{33}$ & 1.2 (0.3--4.1) & 4.0 (1.5--) & 25.3/24 \\
 396 & WR\,20b & 1T & \nodata & 2.5 (1.9--3.4) & 3.0 (1.9--6.4) & \nodata & \nodata & 1.9$\times$10$^{-13}$ & 7.6$\times$10$^{32}$ & 1.0 (--5.3) & 1.6 (--6.4) & 33.6/18 \\
 &         & 2T & (b) & 3.9 (2.2--5.8) & 5.5 (2.8--) & 0.48 (0.30--0.72) & 0.1 & 2.2$\times$10$^{-13}$ & 1.8$\times$10$^{33}$ & 1.2 (0.3--5.0) & 3.4 (0.69--) & 20.3/17 \\
 \hline
 192 & MSP\,188 & 1T & (c) & 0.31 (0.11-0.56) & 3.6 (2.6--6.2) & \nodata & \nodata & 1.2$\times$10$^{-13}$ & 2.9$\times$10$^{32}$ & 6.2 (2.4--11) & 2.1 (0.0--8.0) & 41.7/10 \\
 217 & MSP\,167 & 1T & (d) & 1.4 (0.2--2.8) & 1.4 (0.9--4.7) & \nodata & \nodata & 6.1$\times$10$^{-14}$ & 4.9$\times$10$^{32}$ & 0.9 (0.2--3.1) & 0.5 (--4.7) & 12.5/9 \\
 229 & MSP\,18  & 1T & (e) & 0.63 (0.30--0.99) & 4.1 (2.6--8.9) & \nodata & \nodata & 2.7$\times$10$^{-13}$ & 7.2$\times$10$^{32}$ & \nodata & \nodata & 17.0/18 \\
\enddata
 \tablenotetext{a}{Source numbers follow Table~\ref{tb:t1}.}
 \tablenotetext{b}{``1T'' and ``2T'' indicate the fits by an absorbed one-temperature
 and two-temperature thin-thermal plasma (\texttt{vapec}) model, respectively. The
 amount of extinction (\nh), the plasma temperature (\kt$_{1}$, \kt$_{2}$), X-ray flux
 (\fx) and luminosity (\lx), the abundance of Si and S relative to the cosmic values,
 and the goodness of the fit ($\chi^{2}$/d.o.f.) are derived. The cosmic abundance
 \citep{anders89} is used. For the 2T fits, \kt$_{1}$ and \kt$_{2}$ represent the
 higher and lower temperature component. The \emm\ ratio of the two components is
 given as \emm$_{1}$/\emm$_{2}$.}
 \tablenotetext{c}{The label of figures for the spectra and the best-fit models in
 Figure~\ref{fg:f11}.}
 \tablenotetext{d}{Uncertainties of a 90\% confidence range are given in parentheses.}
 \tablenotetext{e}{Values in the 1.0--8.0~keV range. A distance of 4.2~kpc is assumed
 to derive \lx.}
\end{deluxetable}
\end{turnpage}

\end{document}